\documentclass[%
aps,%
 amsmath,amssymb,
 reprint,%
 superscriptaddress,
nofootinbib,
floatfix
]{revtex4-1}

\usepackage[letterpaper,top=3cm,bottom=2cm,left=1.5cm,right=1.5cm]{geometry} 
\usepackage{graphicx}				

\usepackage[braket, qm]{qcircuit}
\usepackage{braket}
\usepackage{amsfonts}
\usepackage{enumitem}
\usepackage[caption=false]{subfig}
\usepackage{xcolor}

\graphicspath{{figs/}}

\usepackage{xcolor}
\usepackage{soul}

\usepackage{times}

\begin{document}

\title{
Lipkin model on a quantum computer 
}
\author{Michael J.\ Cervia}
\email{cervia@wisc.edu}
\affiliation{%
Department of Physics, University of Wisconsin--Madison,
Madison, Wisconsin 53706, USA
}%
\author{A.\ B.\ Balantekin}
\email{baha@physics.wisc.edu}
\affiliation{%
Department of Physics, University of Wisconsin--Madison,
Madison, Wisconsin 53706, USA
}%
\author{S.\ N.\ Coppersmith}
\email{snc@physics.wisc.edu}
\affiliation{%
Department of Physics, University of Wisconsin--Madison,
Madison, Wisconsin 53706, USA
}%
\affiliation{%
School of Physics, The University of New South Wales,
Sydney, New South Wales 2052, Australia}
\author{Calvin W.\ Johnson}
\email{cjohnson@sdsu.edu}
\affiliation{%
 Department of Physics, San Diego State University, San Diego,
 California 92182-1233, USA
}
\author{Peter\ J.\ Love}
\email{peter.love@tufts.edu}
\affiliation{%
Department of Physics and Astronomy, Tufts University, Medford, Massachusetts 02155, USA}
\affiliation{Computational Science Initiative, Brookhaven National Laboratory, Upton, New York, 11973-5000, USA}
\author{C. Poole}
\email{cpoole2@wisc.edu}
\affiliation{%
Department of Physics, University of Wisconsin--Madison,
Madison, Wisconsin 53706, USA
}%
\author{K.\ Robbins}
\email{Kenneth.Robbins@tufts.edu}
\affiliation{%
Department of Physics and Astronomy, Tufts University, Medford, Massachusetts 02155, USA}
\author{M. Saffman}
\email{msaffman@wisc.edu}
\affiliation{%
Department of Physics, University of Wisconsin--Madison,
Madison, Wisconsin 53706, USA
}%

\date{August 3, 2021}							


\begin{abstract}
  Atomic nuclei are important laboratories for exploring and testing new insights into the universe, such as experiments to directly 
detect dark matter or explore properties of neutrinos. The targets of interest are often 
heavy, complex nuclei that challenge our ability to reliably model them (as well as quantify the uncertainty of those models) with classical computers. Hence there is great interest in applying  quantum computation to nuclear structure 
for these applications. 
As an early step in this direction, especially with regards to the uncertainties in the relevant quantum calculations, we develop circuits to implement variational quantum eigensolver (VQE) algorithms for the Lipkin-Meshkov-Glick model, which is often used in the nuclear physics community as a testbed for many-body methods.
We present quantum circuits for VQE for two and three particles and discuss the construction of circuits for more particles.
Implementing the VQE for a two-particle system on the IBM Quantum Experience, 
we identify initialization and two-qubit gates as the largest sources of error. 
We find that error mitigation procedures reduce the errors in the results significantly, but additional quantum hardware improvements are needed for quantum calculations to be sufficiently accurate to be competitive with the best current classical methods. 

\vspace{1em}\noindent
PhySH: Effective field theory, Particle dark matter, Quantum algorithms, Quantum information with solid state qubits, Many-body techniques, Nuclear many-body theory
\end{abstract}

\maketitle

\section{Introduction}
\label{sec:intro}

Physics today finds itself in a conundrum. On one hand, the standard model of particle physics has been very successful. 
Yet from cosmological observations we are aware of how little we know. The makeup of the universe appears to be dominated by 
nonbaryonic dark matter~\cite{Blumenthal:1984bp,Primack:1988zm,Feng:2010gw,Bertone:2016nfn} and so-called dark energy \cite{Frieman:2008sn}, and even the origin of the matter-antimatter imbalance in the Universe is not fully understood \cite{Dine:2003ax}.  While we understand the basic mechanisms of nucleosynthesis, 
the astrophysical site of a large fraction of heavy elements is still under debate \cite{Kajino:2019abv}. 

Many of the experiments investigating these ongoing mysteries rely upon understanding detailed properties of atomic nuclei, from 
neutrinoless double $\beta$-decay experiments searching for lepton number violation \cite{Dolinski:2019nrj}, to detection of supernova neutrinos \cite{Mirizzi:2015eza}, to the direct detection of dark matter 
\cite{Schumann:2019eaa}. Because many of these experiments place upper limits, it is equally important to quantify 
 the uncertainty in our models of those nuclei~\cite{PhysRevD.87.075014,PhysRevD.87.023512,furnstahl2015recipe,carlsson2016uncertainty,perez2016uncertainty,PhysRevC.98.061301}. 
 
 With the advent of powerful computers and more rigorous techniques, as well as enhanced efforts in uncertainty quantification (UQ), 
 our models of atomic nuclei have improved dramatically in the past two decades. Yet, like physics itself, we paradoxically see all too well 
 the limits of our current computing platforms.  Most of the targets for probing new physics are heavy, 
 complex nuclei such as argon, germanium, or xenon; and uncertainty quantification can require many runs with 
 small variations of parameters~\cite{furnstahl2015recipe,carlsson2016uncertainty,perez2016uncertainty,PhysRevC.98.061301,uqusdb2019}.   For these heavy nuclei, the exponential growth of the Hilbert space dimension 
 makes calculations, especially multiple runs, challenging. 
In this context, the potential of quantum computers is appealing. 
Significant effort is already underway in applying quantum computers to problems with similar features such as quantum chemistry~\cite{Hempel2018,cao2019quantum,googleHF2020}, the structure of atomic nuclei~\cite{Dumitrescu:2018njn,PhysRevD.101.074038}, and the structure of hadrons~\cite{Kreshchuk:2020kcz,mueller2020deeply}.

Useful progress towards implementing on quantum computers standard approximations such as configuration-interaction (CI)~\cite{babbush2017exponentially,Dumitrescu:2018njn} and coupled clusters~\cite{ryabinkin2018qubit,romero2018strategies} has been made. 
However, current quantum computers have much larger errors than classical computers, which must be taken into account when comparing the accuracy of predictions from approximate classical theories and the results of quantum calculations.
Thus, our work here is a necessary first step in understanding the potential applications of quantum computing to nuclear structure needed to  interpret  experiments. 

 To start addressing  quantum computation of models relevant to nuclear targets,
we look at a simplified model of the many-body targets, the Lipkin-Meshkov-Glick (LMG or, colloquially, Lipkin) model~\cite{Lipkin:1964yk} where, because of symmetries, exact solutions are known and can be compared to quantum results.  We present quantum circuits that can
be used to implement variational quantum eigensolver (VQE)~\cite{peruzzo2014variational} algorithms for LMG models with different numbers of particles.
We implement a VQE algorithm for a two-particle LMG model on the International Business Machines Corporation (IBM) Quantum  Experience, a publicly available quantum computer, and identify the main sources of computational errors. We find that errors in measurement and in two-qubit gate operations are critical limitations.  Implementation of error mitigation techniques~\cite{temme2017error,Dumitrescu:2018njn,He:2020udd} provide significant improvement, though the remaining errors are not negligible.  The analysis that we perform on the LMG model illustrates the current limitations of quantum computers and also identifies the improvements needed so that they can be able to provide results superior to those from classical machines.

The paper is organized as follows.
In Sec.~II we use  the direct detection  of dark matter as a case study, and discuss why quantum computers are potentially extremely useful.  We then discuss
how UQ is central to the comparison between quantum and classical computational approaches in a way relevant to experimental progress.
In Sec.~\ref{sec:Lipkin} we define the LMG model and show how its symmetry properties can be exploited to obtain analytic solutions for the ground state that can provide a benchmark for the results of quantum algorithms.
In Sec.~\ref{sec:VQE} we present quantum circuits for determining the ground state of the LMG model using a VQE approach. 
In Sec.~\ref{sec:noise} we implement the algorithm for the smallest nontrivial case on the IBM Quantum Experience.  We discuss the effects of different sources of infidelity in the calculation and their relative contributions to error in these VQE algorithms.
We also explore the effectiveness of error mitigation techniques proposed in Refs.~\cite{temme2017error,Dumitrescu:2018njn,He:2020udd} and show that the improvement in the accuracy of the calculations is substantial.
In Sec.~\ref{sec:dm} we give an example calculation of an observable as a forerunner of the kind of calculation 
one would need for actual applications.
In Sec.~\ref{sec:summary} we summarize our results and sketch further avenues for exploration.

\section{Dark Matter, Nuclear Structure, and Quantum Computing}

Although there are many important applications of nuclear structure physics, here we use the direct experimental 
detection of dark matter as a case study.
Recent observations in astrophysics and cosmology provide strong evidence that a large fraction of our Universe's mass is composed of nonbaryonic dark matter~\cite{Blumenthal:1984bp,Primack:1988zm,Feng:2010gw,Bertone:2016nfn}. The direct detection of particle dark matter by measuring the recoil of nuclei that collide with dark matter particles would not only confirm this picture, it would demonstrate physics beyond the Standard Model \cite{Bertone:2004pz,Jungman:1995df}.


For many years dark matter interactions with baryonic matter were 
simply divided into coupling to the bulk (spin-independent) and coupling to the 
spin of quarks \cite{RevModPhys.71.S197,PhysRevD.31.3059}, but 
recent theoretical developments~\cite{Fitzpatrick:2012ix,Anand:2013yka,Vietze:2014vsa,Fieguth:2018vob,Hoferichter:2020osn} using effective field theory (EFT) techniques, have shown that the interpretation of 
direct detection experiments should be expanded to 
 six (or more if one allows symmetry violation) nucleon-dark matter couplings. 

These theoretical developments have important consequences for experimental design ~\cite{alsum2020effective}.
The target response to scattering of dark matter is computed by folding the single-nucleon reduced density matrix with the one-body matrix elements of operators derived in EFT.
The relative sensitivity of experiments using different nuclear targets can vary by several orders of magnitude under changes in the underlying dark matter-nucleon coupling.

In addition, UQ has begun to be 
implemented into the theory of atomic nuclei~\cite{furnstahl2015recipe,carlsson2016uncertainty,perez2016uncertainty,PhysRevC.98.061301},
based in part upon the realization that correct assessment of any experiment that rests upon models 
needs UQ \cite{PhysRevD.87.075014,PhysRevD.87.023512,uqusdb2019}.

We have very good and predictive theories of nuclear structure, such as but not limited to 
the \textit{no-core shell model} (NCSM), which is an \textit{ab initio} CI method for the wavefunctions of atomic nuclei.  The NCSM and other \textit{ab initio} theories start 
from nucleon-nucleon scattering data and then, without further adjustment of parameters, calculate the structure and spectra of light nuclei~\cite{navratil2000large,barrett2013ab}. While in many aspects such calculations are very successful, the 
application of the NCSM has been limited largely to light nuclides, with mass number $A < 16$. Other 
\textit{ab initio} methods such as coupled clusters can tackle heavier nuclei~\cite{hagen2010ab}, but are mostly limited to 
near closed shells.

Alternatively, one can turn to phenomenological or empirical CI
calculations~\cite{BG77,br88,ca05}. Here one works
in a restricted valence space. The interaction matrix elements, while starting  from `realistic' 
forces similar to those used in the NCSM, are adjusted to fit many-body spectra.  Thus, phenomenological
CI calculations have less rigorous foundations, when compared to the NCSM, and yet a greater range of 
applicability.   (There are efforts 
to connect \textit{ab initio} methods to phenomenological-like spaces with greater rigor and predictive power~\cite{doi:10.1146/annurev-nucl-101917-021120}, but those are still in development.)
This comparison is particularly true with regard to medium- and heavy-mass nuclides of interest to the current generation of dark matter detectors.

Our challenge is that  CI calculations~\cite{ca05} needed for dark matter calculations   \cite{PhysRevD.40.2131,PhysRevD.48.5519,PhysRevD.93.095012,PhysRevD.87.075014,PhysRevD.87.023512,PhysRevD.86.103511,PhysRevD.88.083516,Vietze:2014vsa,PhysRevD.88.115014,PhysRevD.95.103011} suffer from the exponential growth in the cost of storing the wavefunction classically.
The largest 
CI calculations  to date work in a basis space of dimension
of the order $10^{10}$. 
However, $^{40}$Ar, a key target in many experiments, if one works in a nuclear valence 
space of $1s$-$0d$-$1p$-$0f$ orbits, has a $M$-scheme (fixed-$J_z$) basis dimension of nearly 10$^{15}$.
Typically, one restricts excitations from the $1s$-$0d$ orbits into the $1p$-$0f$ orbits~\cite{PhysRevD.99.055031}, but the results depend upon the specific truncation. 
For Xe isotopes, phenomenological calculations are generally in the restricted $0g_{7/2}$-$2s$-$1d$-$0h_{11/2}$ space.  The most common isotope, $^{132}$Xe (with a natural abundance of $29.9\%$), requires a 
$M$-scheme dimension of only $3.7\times10^7$, which can be calculated on a powerful laptop. 
The next most common isotope, $^{129}$Xe, has a basis dimension of $3\times10^9$, which can only be 
calculated on a supercomputer. 
$^{128}$Xe ($1.9\%$) has a basis dimension of $9.3\times10^9$, and $^{124}$Xe, rare yet also of 
interest to neutrinoless double-electron capture decay, has a $M$-scheme basis dimension
of $1.86\times10^{11}$, beyond the reach of current supercomputers. 


For phenomenological calculations, UQ is both empirical and time-consuming. 
 One varies the interaction parameters, of which there can be dozens or even hundreds, and 
recomputes the energies and other observables, in order to build up a model of the multi-dimensional 
error surface~\cite{PhysRevC.98.061301,uqusdb2019}. While this can be done in small model spaces where 
one can compute hundreds of observables in a few minutes, in larger spaces, where calculations of a 
single nuclide can take hundreds of CPU hours, such UQ analyses are daunting. Here is one example where 
even near-term quantum computers could be helpful in dramatically speeding up the many
large calculations needed for UQ.

Quantum simulators have the potential to transform our ability to understand the performance of 
experiments, 
based on the ability of quantum simulators to calculate the properties of ground states of fermionic Hamiltonians much more efficiently than currently known classical algorithms. 
Performing these calculations using quantum computers 
is potentially advantageous because
simulation of fermions is efficient~\cite{lloddse} and does not suffer~\cite{PhysRevD.101.074510} from the ``sign problem" that places severe limits on system sizes and/or temperatures achievable in fermionic calculations done using classical computers~\cite{Loh:1990zz}.
A key question is how much reliable information can be obtained from current noisy quantum
computers.
To address this question, here we investigate a simplified model of targets
that has symmetry properties enabling classical computers to
determine the ground states of large systems; indeed, substantial
analytic results are also available.



\section{Lipkin-Meshkov-Glick Model}
\label{sec:Lipkin}

To assess the validity of various quantum computational techniques, we calculate the ground state wave function of the LMG model~\cite{Lipkin:1964yk}. The LMG model is widely used as a testbed 
for approximations in many-body physics, for example, time-dependent Hartree-Fock \cite{krieger1977comparison}, 
time-dependent coupled-clusters \cite{PhysRevC.18.2380,wahlen2017merging}, the random phase approximation \cite{PhysRevC.64.017303}, 
generator coordinate methods \cite{PhysRevC.74.024311}, and density functional theory \cite{PhysRevC.79.014301,PhysRevC.78.064310}, a list which barely scratches  the surface. It therefore strikes us 
as sensible to also use the LMG  model as an early implementation of quantum computation.

In the LMG model~\cite{Lipkin:1964yk}, $N$ fermions are distributed among two levels with $N$-fold degeneracy and an energy separation of $\epsilon$. Defining $c^{\dagger}_{\sigma p}$ and $c_{\sigma p}$ as the creation and annihilation operators of the fermion in the state $p$ of level $\sigma ~(=\pm 1)$, we  write the Hamiltonian of the system as 
\begin{align}
\label{LMGH}
{\tilde H} =& \frac{1}{2} \epsilon \sum_{\sigma,p} \sigma c^{\dagger}_{\sigma p} c_{\sigma p} \nonumber \\
    &+ \frac{1}{2} {\tilde V} \sum_{\sigma, p, p'} c^{\dagger}_{\sigma p} c^{\dagger}_{\sigma p'} c_{- \sigma p'} 
c_{- \sigma p} . 
\end{align}
A term that scatters one fermion to the upper level and a second fermion to the lower level can also be added to this Hamiltonian, but such a term yields a constant in the SU(2) subspaces described below. Introducing the quasi-spin operators 
\begin{align}
J_+ &= \sum_p c^{\dagger}_{+1 p} c_{-1 p} = (J_-)^{\dagger}, 
\label{eq:Jpm} \\ 
J_0 &= \frac{1}{2}  \sum_{\sigma, p} \sigma c^{\dagger}_{\sigma p} c_{\sigma p} ,
\label{eq:Jz}
\end{align}
which span a SU(2) algebra, the LMG Hamiltonian can be rewritten as 
\begin{equation}
\label{eq:LMGSU2}
{\tilde H} = \epsilon J_0 + \frac{1}{2} {\tilde V} \left( J_+^2 + J_-^2 \right) .
\end{equation}
One can calculate the expectation value of the Hamiltonian in Eq.~\eqref{eq:LMGSU2} in the total quasispin basis. 
For a given $N$, a matrix representing this operator has dimension $2^N$, but it consists of blocks of $(2j+1)\times(2j+1)$ matrices with SU(2) labels 
corresponding to different $j$ values obtained by adding $N$ SU(2) doublets. In the rest of this paper, we work with the dimensionless Hamiltonian $H\equiv {\tilde H}/\epsilon$ with $V\equiv {\tilde V}/\epsilon$. Also we will only consider the multiplet with $j=N/2$ containing the unperturbed ground state. 

It is especially convenient to write the Hamiltonian in the qubit basis. For $N$ particles, the total quasispin is given by 
\begin{equation}
{\mathbf J} = \sum_{p=1}^N {\mathbf J}^{(p)} 
\end{equation}
where each ${\mathbf J}^{(p)}$ is in the $j=1/2$ representation. Hence, the Hamiltonian of Eq.~\eqref{eq:LMGSU2} becomes 
\begin{equation}
\label{eq:spinH}
H = \sum_{p=1}^N J_0^{(p)} + V  \sum_{\substack{p,q=1\\q\neq p}}^N \left( J_+^{(p)} J_+^{(q)} + J_-^{(q)} J_-^{(p)} \right) .
\end{equation}
That is, the Hamiltonian matrix elements are: the sum of the $J_z$ values of the qubits ($\pm 1/2$) along the diagonal entries, the quantity $V$ when two qubits can be flipped, or zero otherwise. 

\section{Variational Quantum Eigensolver}
\label{sec:VQE}

In this section, we outline specific VQE algorithms for computation of the ground state of LMG models for generic values of $V$ and with fixed values of $N$. We introduce the algorithm with the example of $N=2$, for which we carry out calculations on quantum hardware and in noise simulations in later sections, and then present two directions for generalization of this method.



\subsection{$N=2$}
\label{sec:VQE2}

We set up our algorithm by defining a dictionary basis to correspond to the quasi-spin basis; for example with $N=2$:
$\{\ket{00},\ket{01},\ket{10},\ket{11}\}$ $=$ $\{\ket{\uparrow\uparrow},\ket{\uparrow\downarrow},\ket{\downarrow\uparrow},\ket{\downarrow\downarrow}\}$.
In this basis the Hamiltonian is represented by 
\begin{align}
H &\doteq \frac{1}{2}\left( \sigma_3 \otimes 1 + 1\otimes \sigma_3 \right) + \frac{V}{2} \left( \sigma_1 \otimes \sigma_1 - \sigma_2 \otimes \sigma_2 \right)  \nonumber \\
&=
\begin{pmatrix}
+1 & 0 & 0 & V  \\
 0 & 0 & 0 & 0 \\
 0 & 0 & 0 & 0 \\
 V & 0 & 0 & -1
\end{pmatrix}. 
\label{eq:2qubitHamiltonian}
\end{align} 
Here, $\sigma_1$, $\sigma_2$, and $\sigma_3$ are the Pauli matrices. It is then straightforward to diagonalize this Hamiltonian to obtain the eigenvalues $0$ (with multiplicity 2) and $\pm \sqrt{1+V^2}$. The normalized ground state with energy $E_\mathrm{gnd}=-\sqrt{1+V^2}$ is given by 
\begin{align}
    \ket{\Psi} = \frac{V\ket{\uparrow\uparrow}- (1+\sqrt{1+V^2}) \ket{\downarrow\downarrow}}{\sqrt{2V^2+2+2\sqrt{1+V^2}}}
    \label{eq:exactgroundstate}
\end{align}
Here we consider a trial state that is a real superposition of the two states with total quasi-spin $j=1$ and $|m|=1$:
\begin{equation}
\label{eq:N=2trial}
\ket{\psi(\theta)} = \sin(\theta) \ket{\uparrow\uparrow} -\cos(\theta)\ket{\downarrow\downarrow},
\end{equation}
defining a single variational parameter $\theta$ that can be optimized to minimize the value of $\bar{H}(\theta)\equiv\braket{\psi(\theta)|H|\psi(\theta)}$. 
The state at which
 $\tan(\theta) = V/(1+\sqrt{1+V^2})$ is the exact ground state, and so we restrict consideration of $\theta$ to the domain $[0,\pi/2)$.
In our VQE we optimize the value of $\theta$ by minimizing the expectation
value of the energy $\bar{H}(\theta)$ evaluated on a quantum computer. 

\begin{figure}[h]
\includegraphics[width=0.475\textwidth]{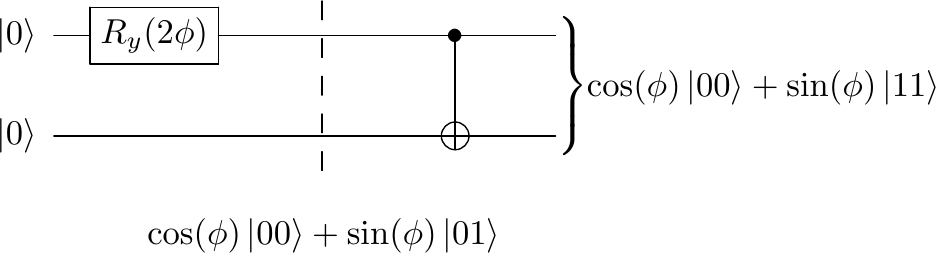}
\caption{ \label{fig:circuit} 
A quantum circuit to prepare the two-qubit VQE trial state $\ket{\psi(\theta)}$ given by Eq.~\eqref{eq:N=2trial} with $\phi=\theta-\pi/2$ on a noiseless quantum device.
At the dashed line, the intermediate state is $\cos(\phi) \ket{00}+\sin(\phi) \ket{01}$.
}
\end{figure}

The state given by Eq.~\eqref{eq:N=2trial} can be prepared from an initial state $\ket{00}$ by applying a one-qubit rotation about the $y$ axis of the Bloch sphere of a first qubit, written as
\begin{equation}
    R_y(2\phi) = \mathrm{exp} \bigg\{-\frac{i}{2}(2 \phi) Y\bigg\},
    \label{eq:Ry}
\end{equation}
with $\phi =\theta-\pi/2$, followed by a CNOT gate using the first qubit as the control and a second qubit as the target:
\begin{equation}
    \mathrm{CNOT}_{c=1,t=2} = 
\begin{pmatrix}
1 & 0 & 0 & 0 \\
0 & 1 & 0 & 0 \\
0 & 0 & 0 & 1 \\
0 & 0 & 1 & 0
\end{pmatrix}
    \label{eq:CNOT}
\end{equation}
Here, $X$, $Y$, and $Z$ are the Pauli gates, $X=\begin{pmatrix} 0 & 1 \\ 1 & 0 \end{pmatrix}$, $Y=\begin{pmatrix} 0 & -i \\ i & 0 \end{pmatrix}$, $Z=\begin{pmatrix} 1 & 0 \\ 0 & -1 \end{pmatrix}$. 
The quantum circuit for state preparation is summarized in Fig.~\ref{fig:circuit}.

The measurements on this quantum circuit needed to compute $\bar{H}$ are simultaneous measurements in the $X$ basis for both qubits and in the $Y$ basis for both qubits as well as measurements of each qubit in the $Z$ basis. These measurements yield estimates of the expectation values $\braket{XX}$, $\braket{YY}$, $\braket{ZI}$, and $\braket{IZ}$, and thus determine $\bar{H}$ from Eq.~\eqref{eq:spinH}.

This procedure of writing the ground state wave function for $N$ particles as a real superposition controlled by a single trial parameter on a $N$-qubit device may be generalized for higher values of $N$, as we demonstrate in Sec.~\ref{sec:VQE34}. 
However, first, let us briefly comment on the dimensionality of the ground-state Hilbert space within the LMG model. 
In general, the ground state will be a superposition of the state with all quasispins down and all states with any number pairs of quasi-spins flipped up. This parity is a symmetry that we can to exploit to reduce the cost of preparing a variational state and requiring only a single variational parameter.
For this reason, we excluded both of the states with $m=0$ above, and so this dimension was simply 2 for the case $N=2$. For general $N$, this dimension will then be $2^{N-1}$. 

\subsection{Variational states for $N=3,~4$}
\label{sec:VQE34}

In Sec.~\ref{sec:VQE2}, we presented a quantum circuit to obtain a trial state with a single parameter to estimate the ground state of a LMG model for $N=2$. Here, we generalize this approach to obtain variational wave functions for $N=3$ and $N=4$.

\subsubsection{Trial state for $N=3$}
\label{sec:3trial}

For $N=3$, using the usual basis 
\begin{align*}
    &\{\ket{000},\ket{001},\ket{010},\ket{011},\cdots,\ket{111}\} \\
    =
    &\{\ket{\uparrow\uparrow\uparrow},\ket{\uparrow \uparrow \downarrow},\ket{\uparrow \downarrow\uparrow},\ket{\uparrow\downarrow\downarrow},
    \cdots,\ket{\downarrow\downarrow\downarrow}\},
\end{align*}
the Hamiltonian is represented by 
\begin{eqnarray}
H &\doteq& \frac{1}{2}( \sigma_3 \otimes 1 \otimes 1  + 1 \otimes \sigma_3 \otimes 1 + 1 \otimes 1 \otimes \sigma_3 ) \nonumber \\ 
 &+& \frac{V}{2} (1 \otimes \sigma_1 \otimes \sigma_1 + \sigma_1 \otimes \sigma_1 \otimes 1 + \sigma_1 \otimes 1 \otimes \sigma_1 \nonumber \\
&-& 1 \otimes \sigma_2 \otimes \sigma_2 - \sigma_2 \otimes \sigma_2 \otimes 1 - \sigma_2 \otimes 1 \otimes \sigma_2)
 \nonumber \\
&=&
\begin{pmatrix}
\frac{3}{2}  &  0 & 0 &V & 0 & V & V & 0   \\
0  & \frac{1}{2}   & 0 & 0 & 0 & 0 & 0 & V \\
0  & 0  &  \frac{1}{2} & 0 & 0 & 0 & 0 & V  \\
V & 0 & 0 & - \frac{1}{2} & 0 & 0 & 0 & 0 \\
0 & 0 & 0 & 0 & \frac{1}{2} & 0 & 0 & V \\
V & 0 & 0 & 0 & 0 & -\frac{1}{2} & 0 & 0  \\
V & 0 & 0 & 0 & 0 & 0 & -\frac{1}{2} & 0  \\
0 & V & V & 0 & V & 0 & 0 & -\frac{3}{2} 
\end{pmatrix}.
\label{eq:3qubitHamiltonian}
\end{eqnarray}
Similar to Eq.~\eqref{eq:N=2trial} for $N=2$, an appropriate variational ansatz  for the ground state of $N=3$  is 
\begin{align}
 \ket{\psi(\theta)} = &\cos(\theta)\ket{\downarrow\downarrow\downarrow} 
 \nonumber \\
&- \frac{1}{\sqrt{3}} \sin(\theta) \big(\! \ket{\uparrow\uparrow\downarrow} + \ket{\uparrow\downarrow\uparrow} + \ket{\downarrow\uparrow\uparrow} \!\big),
\label{eq:N=3trial}
\end{align}
where $\theta$ is the variational parameter.  This wave function is the ground state when $\sqrt{3}\cot(\theta)=V/(1+\sqrt{1+3V^2})$ with energy $-1/2-\sqrt{1+3V^2}$, and so we restrict consideration of $\theta$ to the domain $[0,\pi/2)$.

\begin{figure}
\includegraphics[width=0.4\textwidth]{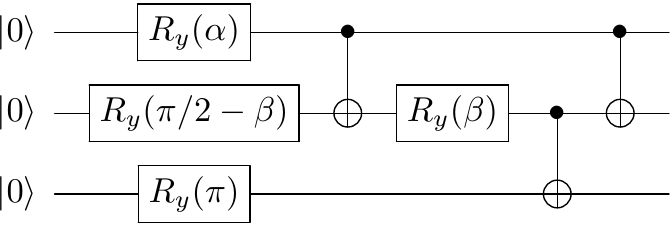}
\caption{ \label{fig:circuit3} A quantum circuit to prepare the three-qubit VQE trial state $\ket{\psi}$ given by Eq.~\eqref{eq:N=3trial}, using variables defined in Eqs.~\eqref{eq:alphaparam}--\eqref{eq:betaparam}. }
\end{figure}

The three-qubit preparation circuit  is shown in Fig.~\ref{fig:circuit3}, written with two auxiliary angles $\alpha$ and $\beta$ defined by
\begin{align}
\alpha &\equiv 2\arccos\bigg(-\sqrt{\frac{2}{3}}\sin\theta\bigg), \label{eq:alphaparam} \\
\beta &\equiv -\frac{\pi}{4} - \arctan\bigg(\frac{\tan\theta}{\sqrt{3}}\bigg). \label{eq:betaparam}
\end{align}

\subsubsection{Trial state for $N=4$}
\label{sec:4trial}

We can prescribe similarly a four-qubit trial state for the LMG model with $N=4$. Here, the ground state belongs to the $j=2$ representation and has the energy $-2 \sqrt{1+ 3V^2}$.
The unnormalized ground state wave function with this energy is 
\begin{align}
\ket{\Psi} = &\ket{\downarrow\downarrow\downarrow\downarrow}+A\ket{\uparrow\uparrow\uparrow\uparrow}  \nonumber \\
- B\big(\!&\ket{\uparrow\uparrow\downarrow\downarrow} + \ket{\downarrow\downarrow\uparrow\uparrow} + \ket{\downarrow\uparrow\downarrow\uparrow} \nonumber \\
\phantom{+B}+ &\ket{\downarrow\uparrow\uparrow\downarrow} + \ket{\uparrow\downarrow\downarrow\uparrow} + \ket{\uparrow\downarrow\uparrow\downarrow} \!\big)
\label{eq:4state}
\end{align} 
where 
\begin{align}
A & = 1 - 2 \frac{\sqrt{1+3V^2}-1}{3V^2}, \label{eq:Aparam} \\
B & = \frac{\sqrt{1+3V^2}-1}{3V} . \label{eq:Bparam}
\end{align}
After normalization one finds that the coefficients of $\ket{0000}$ and $\ket{1111}$ in Eq.~\eqref{eq:4state} sum to 1. Therefore, we propose for $N=4$ a normalized trial state 
\begin{align}
\ket{\psi(\theta)} =& \cos^2\theta \ket{\downarrow\downarrow\downarrow\downarrow} + \sin^2\theta \ket{\uparrow\uparrow\uparrow\uparrow} \nonumber \\
 -&\frac{1}{\sqrt{12}} \sin2\theta \big(\ket{\uparrow\uparrow\downarrow\downarrow} + \ket{\downarrow\downarrow\uparrow\uparrow} + \ket{\downarrow\uparrow\downarrow\uparrow} \nonumber \\
 \phantom{+}&\phantom{\frac{1}{\sqrt{12}}\sin2} + \ket{\downarrow\uparrow\uparrow\downarrow} + \ket{\uparrow\downarrow\downarrow\uparrow} + \ket{\uparrow\downarrow\uparrow\downarrow} \!\big) 
 \label{eq:N=4trial}
\end{align} 
with one variational parameter $\theta$. 
The true ground state of the system is of this form with $\theta$ satisfying $\tan(2\theta)=\sqrt{12}B/(1-A)$, and so we restrict consideration of $\theta$ to the domain $[0,\pi/2)$.

We can continue this process for $N > 4$, establishing a trial wave function depending on one variational parameter for each $N$. A common feature of these wave functions is that they all have definite parity. 
While a $N$-qubit state of definite parity can be constructed from an associated $(N-1)$-qubit state using an additional $N-1$ {\small CNOT} gates, the
associated $(N-1)$-qubit state will not in general have symmetries to exploit, and so a generic quantum state preparation routine is necessary to produce it. Using the quantum state preparation method of Ref.~\cite{quantumstateprep} to prepare the associated $(N-1)$-qubit state, the {\small CNOT} cost of preparing an arbitrary $N$-qubit state of definite parity is $(115/192)2^N - (7/4) 2^{N/2}+N+2/3$ for even $N$ and $(23/48)2^N-2^{(N+1)/2}+N+2/3$ for odd $N$.

The next subsection presents a method for constructing quantum circuits that
generate a $N \ge 4$ particle variational state in a bosonic representation.
Moreover, the method can be used to construct quantum circuits for generating the appropriate variational state for any $N$.

%


\subsection{VQE circuits for $N \ge 4$}
\label{sec:VQEN}




The LMG Hamiltonian can be rewritten in terms of bosonic operators acting on two bosonic modes~\cite{ORTIZ2005421}:
\begin{equation}\label{bham}
H=\frac{n_b-n_a}{2}+\frac{V}{2N}\left(b^\dag b^\dag aa+ a^\dag a^\dag bb\right),
\end{equation}
where $a^\dagger$ and $a$ ($b^\dagger$ and $b$) are the creation and annihilation operators for a boson in the mode $a$ ($b$),
and $n_a,n_b$ are their number operators. For the bosonic representation, the number of particles $N=2j$ is equal to the particle number of the fermionic representation~\cite{LermaH.:2013cla}.

Since the LMG model is exactly solvable any eigenstate of the LMG Hamiltonian $\ket{\psi_{a,b}}$ can be written as the operator~\cite{ORTIZ2005421}
\begin{equation}
   \prod^M_{\ell=1}\left(\frac{(a^\dag)^2}{E_\ell-1}+\frac{(b^\dag)^2}{E_\ell+1}\right)
   \label{eq:ego}
\end{equation}
acting on bosonic fiducial state $\ket{\nu_a,\nu_b}$. The integer $M$ is related to $N$ and $\nu_a,\nu_b$ by 
$N=2M+\nu_a+\nu_b$, where $\nu_a,\nu_b$ are initially restricted to be $0$ or $1$. For nonzero $V$ the spectral parameters $E_\ell$ are real numbers obtained by solving the Bethe ansatz equations~\cite{ORTIZ2005421,LermaH.:2013cla}
\begin{multline}\label{eq:beqs}
    1+\frac{V}{N\left(E_\ell^2-1\right)}\bigg[\left(\nu_a-\nu_b\right)\left(1+E_\ell^2\right)\\
    +2E_\ell\left(1+\nu_a+\nu_b\right)\bigg]\\
    +2\frac{V}{N}\sum^M_{\substack{n=1\\n\neq\ell}}\frac{1+E_\ell E_n}{E_\ell-E_n}=0.
\end{multline}

The two bosonic modes can be encoded in qubits up to a cutoff in occupation number by standard techniques~\cite{somma2005quantum}. The product nature of the exact solution Eq.~\eqref{eq:ego} lends itself naturally to the definition of a quantum circuit for preparation of the exact eigenstates for any number of particles. The general LMG eigenstate generating circuit, explored in more detail in Ref.~\cite{robbins2021benchmarking}, has a depth of $\mathcal{O}(\log_2 N)$ and uses $\mathcal{O}(N)$ gates which act on $\mathcal{O}(N/2)$ qubits.

\section{Results of quantum calculations using the IBM Quantum Experience}
\label{sec:noise}
In this section we implement the VQE calculation for a LMG model with $N=2$ on a quantum computer and characterize the importance of different decoherence errors.
Quantifying these errors yields some insight into how much the performance of quantum computers needs to be improved for quantum calculations to yield results that are more accurate than those obtained using approximate classical methods.
For the calculations reported here, we fix $V=1$, where both one-qubit and two-qubit operators contribute at comparable scales to $\bar{H}$.

We use the open source Quantum Information Science Kit (QISKit, or Qiskit)~\cite{qiskitAbbrev} and run the quantum algorithms on the ibmq\_16\_melbourne, the device with the largest number of qubits that is made publicly available by IBM through their Quantum Experience program. We refer to this device as the ``Melbourne processor." We also perform calculations on IBM Quantum (Q) Experience's open quantum assembly language (QASM)~\cite{ibmGlossary} Simulator (or ``qasm\_simulator'')~\cite{OpenQASM} and investigate the effects of different decoherence mechanisms, which helps to identify the physical improvements that would yield the largest increases in the calculational accuracy.

\subsection{Error characterization}
\label{sec:errors}

Qiskit, which is the software interface for the IBM Quantum Experience, provides a mechanism for including errors that are obtained by fitting the results of a calibration run of the quantum device to a  combination of errors of specific types~\cite{GeneralNoise}:
(1) single-qubit thermal relaxation errors,
(2) single-qubit depolarizing errors,
(3) two-qubit gate depolarizing errors,
(4) single-qubit thermal relaxation errors of both qubits in a two-qubit gate, and
(5) single-qubit readout errors. 

The relaxation errors are parameterized using the relaxation time $T_1$; the fidelity of single-qubit gates is determined by the product of $T_1$ and the qubit frequency, while the relaxation-induced infidelity of the two-qubit gates is determined by the ratio of $T_1$ and the duration of the gate.
Depolarization errors of the single qubit gates are parameterized by dephasing times $T_2$, where again the relevant parameter is the ratio of $T_2$ to the gate duration.
The two-qubit depolarization errors quantify errors that occur in addition to the relaxation errors of the individual qubits during the gate duration. For the readout errors,
measurement errors for the states $\ket{0}$ and $\ket{1}$ are typically different, and the Qiskit error class provides two readout errors, $P(0|1)$ and $P(1|0)$,
as described in Ref.~\cite{ReadOutNoise}.

The parameters that quantify these error sources as obtained from the calibration data of IBM Q backends~\cite{BackendProperties} are presented in Table~\ref{parameters}.
The U2 gate listed in the table is a single-qubit rotation about the $x+z$ axis of the Bloch sphere. The error parameters for U1 gates (rotations about the $z$ axis of the Bloch sphere) and U3 gates (generic single-qubit rotations with three Euler angles) are not listed in the table, as U1 gates are implemented classically via post-processing~\cite{U1gate} and U3 gates are implemented by composing U1 and U2 gates~\cite{U3gate}.
We note that the error parameters are different for different qubits.  

Additionally, we note that the Qiskit-provided ``noise models'' are fits of randomized benchmarking data~\cite{QiskitBenchmarking} to simplified approximate descriptions of IBM Q device errors, as opposed to a comprehensive description of all modes of error in a noisy quantum device~\cite{NoiseModelApprox}, and that the calibration data are obtained from a daily measurement protocol of the device backend and may vary over the course of the day. 

\begin{table}
\caption{\label{parameters}Values of error parameters for qubits 1 and 2 in the Melbourne processor.\\}
\begin{ruledtabular}
\begin{tabular}{lcr}
 	&	\multicolumn{ 2}{c}{Qubit} \\ \cline{2-3}
{Parameter} & \multicolumn{1}{c}{1} & \multicolumn{1}{c}{2}	\\ 
\colrule
$T_1$ ($\mu$s)		&	$48.0$	&	$50.9$	\\
$T_2$ ($\mu$s)		&	$60.2$	&	$48.1$	\\
Qubit frequency (GHz)     &	$5.1$	&	$5.2$	\\
U2 gate length (ns) &   $53$    &   $53$    \\
{\small CNOT} gate length (ns)&  $740$   &   $690$   \\
U2 gate error       &	$0.0005$&	$0.0013$\\
{\small CNOT} gate error     &	$0.0255$&	$0.0255$\\
Readout error $P(0|1)$& $0.005$ &   $0.019$ \\
Readout error $P(1|0)$& $0.046$ &   $0.101$ \\
\end{tabular}
\end{ruledtabular}
\end{table}

We report the errors obtained in a classical simulation of the processor when each of these types of error is either included or excluded and compare these errors to the results obtained using the Melbourne quantum processor.
This comparison enables us to identify the error sources that are currently limiting the performance, for which mitigation would improve the accuracy the most.

\subsection{Results}
\label{sec:results}

To probe the degree of contributions from the different sources of infidelity to the errors in the results for the LMG model, we calculate the difference between the energy obtained as a result of our VQE algorithm, $\min_{\theta}\bar{H}(\theta)$, and the ground state energy, $E_\mathrm{gnd}=-\sqrt{2}$ for $V=1$
for a set of runs on an IBM-supplied classical simulator of the quantum computer that incorporates different subsets of the errors in Table~\ref{parameters}.
We compute the values of 
$\min_\theta\bar{H}(\theta)$
for each of $2^5$ configurations of the
error terms (all combinations in which
each error type is either ``off'' or ``on,'' with magnitude equal to that obtained by fitting the results of the calibration runs)
and compare the results to the exact
ground state energy.

\begin{figure}
\centering
\includegraphics[width=0.49\textwidth,page=1]{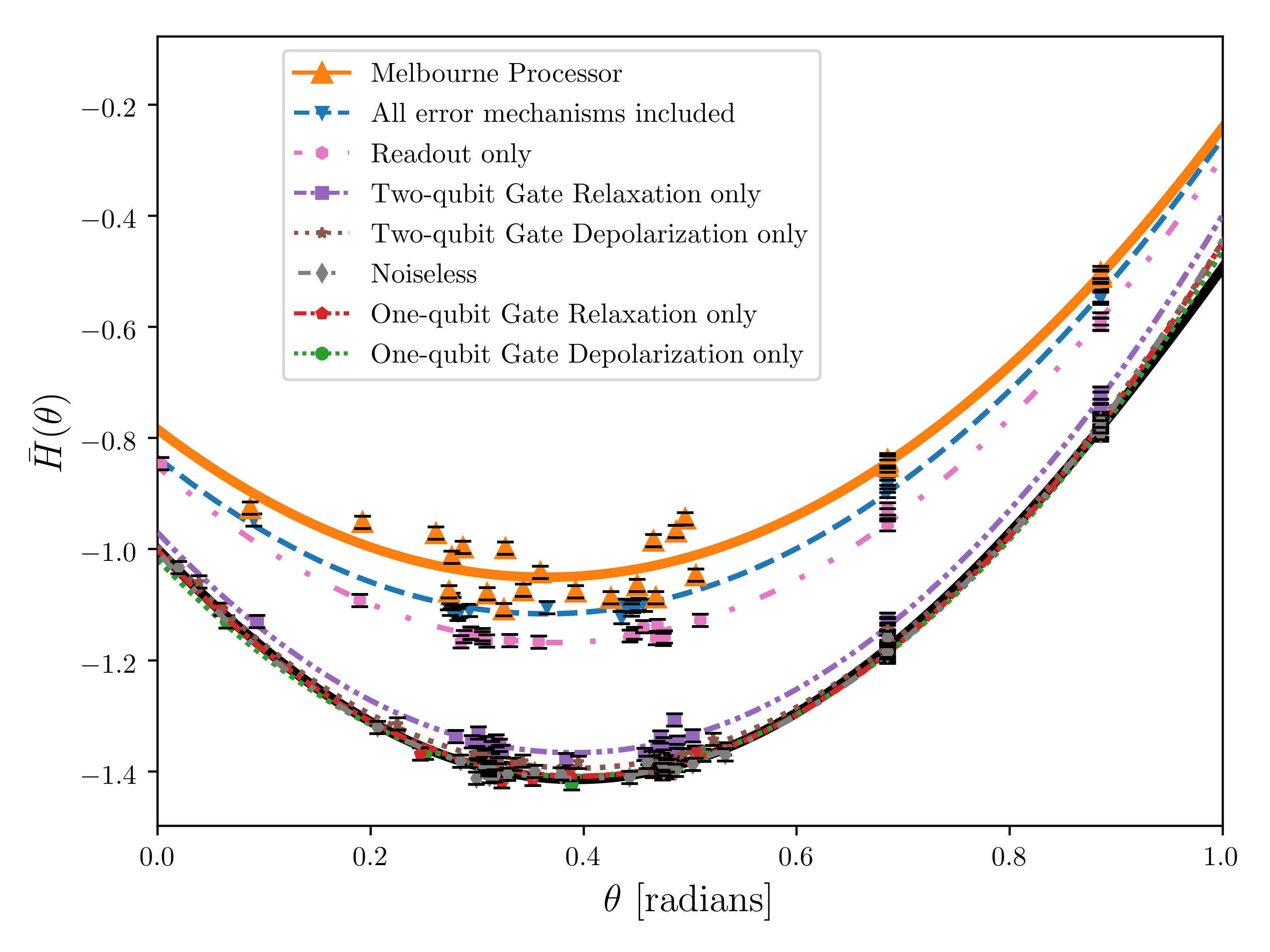} 
\caption{
    Energy of variational state as a function of the VQE trial parameter $\theta$.  The results obtained using the Melbourne processor are compared to the exact result as well as to the results of classical simulations of the quantum processor that incorporate one of the five error mechanisms listed in Table~\ref{parameters}. The lines in the figure are parabolic fits for the Melbourne and simulation results, while the solid black line shows the exact value of $\bar{H}$ obtained analytically using Eq.~\eqref{eq:exactgroundstate}. Error bars on each data point show the statistical errors $1/\sqrt{n}$, where $n=8192$ is the number of runs over which the result for each model is averaged.
} \label{fig:singlenoise}
\end{figure}

Figure \ref{fig:singlenoise} shows the energy as a function of the variational parameter $\theta$ with no errors, with all the sources of error in Table~\ref{parameters} included, and with each error type included individually. 
Based on the deviation of the measured energy of the variational state with the exact result, it appears that readout errors dominate the overall error, with two-qubit gate errors the second-largest source of overall error.

We note that the error of the classical simulation is slightly smaller than that of the quantum processor.
It is entirely possible that this discrepancy arises because of the use of an approximate error model and/or because of drift leading to slightly degraded performance over the course of a day, as mentioned above.
It is also possible that the quantum processor has significant initialization errors that are not included in the IBM-provided ``noise model.''

\begin{figure}[htbp]
\centering
\includegraphics[width=0.49\textwidth, page=1]{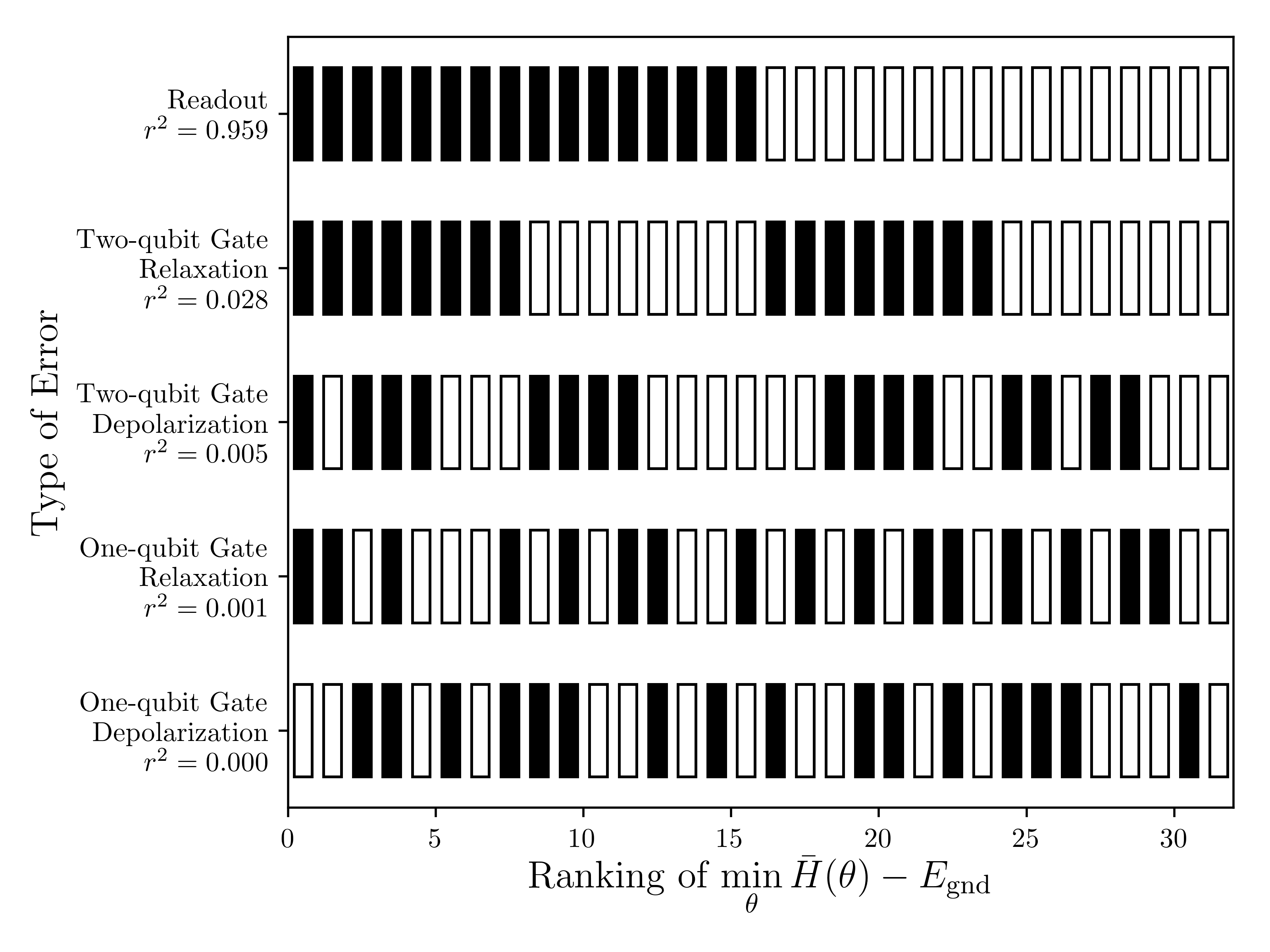} 
\caption{
    We consider $2^5=32$ different combinations of the following sources of computational error (referred to as ``noise models'' in the IBM Qiskit documentation): readout errors, relaxation during two-qubit gates, depolarization errors during two-qubit gates, relaxation errors during one-qubit gates, and depolarization errors during two-qubit gates.  Each error is either off or on, with magnitude given by the calibration run of the quantum processor.
    The rectangles in the plots represent the results of calculations of 32 different combinations of error terms, ordered by increasing variational energy from left to right. Each rectangle is solid (unfilled) if the error type of its row is excluded from (included in) the classical simulation of the quantum algorithm. 
    From this plot one can see that the readout error is dominant, since all of the 16 lowest energies are calculations in which there are no readout errors.
    The two-qubit gate relaxation error is the second most important, since for a given setting of the readout error (either off or on), all eight of the lowest energy results are obtained when the gate relaxation error is off.
    The importance of an error type can be quantified using the correlation coefficient $r^2$ between the value of the binary function of whether the error is on or off and the value of the variational energy.
} \label{fig:modelsOnOff}
\end{figure}

Figure \ref{fig:modelsOnOff} shows a different method for assessing the relative importance of the different errors listed in Table~\ref{parameters}. We track the correlation of each type's setting (whether it is off or on) with the ranking of its corresponding result $\min_\theta\bar{H}(\theta)-E_\mathrm{gnd}$ amongst all other noise models. The figure summarizes these results and also reports the correlation coefficients for each source of computational error. 
This analysis confirms that readout errors are the most significant with two-qubit gate relaxation errors being the second most important and two-qubit gate depolarization errors the third most important.  The effects of one-qubit errors on the variational energies are much smaller than those of the readout errors and of the two-qubit gates.

\subsection{Implementing error mitigation}
\label{subsec:error_mitigation}

In this subsection we investigate the performance of error mitigation procedures for the errors arising from readout and from {\small CNOT} gates for the calculations of the energy of the Lipkin model with $N=2$.  The readout errors are mitigated by using features from the Qiskit library~\cite{QiskitErrorMitigate} in which the measured readout error is used to generate and invert a matrix to obtain the relevant correction.  The two-qubit gate depolarization errors are mitigated using zero-noise extrapolation (ZNE) as described in Ref.~\cite{He:2020udd}.

We first discuss the procedure to mitigate the measurement errors as implemented in the Qiskit library~\cite{QiskitErrorMitigate}.
First, the measurement errors are calibrated.  For our situation with two qubits, one measures the expectation values $\braket{Z_1}$, $\braket{Z_2}$, and $\braket{Z_1Z_2}$ of the states $\{\ket{00}$, $\ket{01}$, $\ket{10}$, $\ket{11}\}$.
In the absence of readout error, each of these measurements would yield the relevant dictionary basis element with unit probability.
In the presence of readout error, the results can be described using a $4\times4$ matrix
\begin{equation}
    M_{ij}=P(i|j),
    \label{eq:calibrationMatrix} 
\end{equation}
where $P(i|j)$ is the probability that the result $i$ is obtained when one measures the basis element $\ket{j}$.
Finally, the probability distribution of measured results from the quantum circuit, $P(i)$, is corrected by writing the distribution as a $2^N=4$-dimensional vector $\vec{P}$ and applying the inverse of the matrix $M$ to a obtain a probability vector with mitigated readout error
\begin{equation}
    \vec{P}' = M^{-1}\vec{P}.
    \label{eq:mitigatedProbability}
\end{equation}

\begin{figure}
\includegraphics[width=0.49\textwidth]{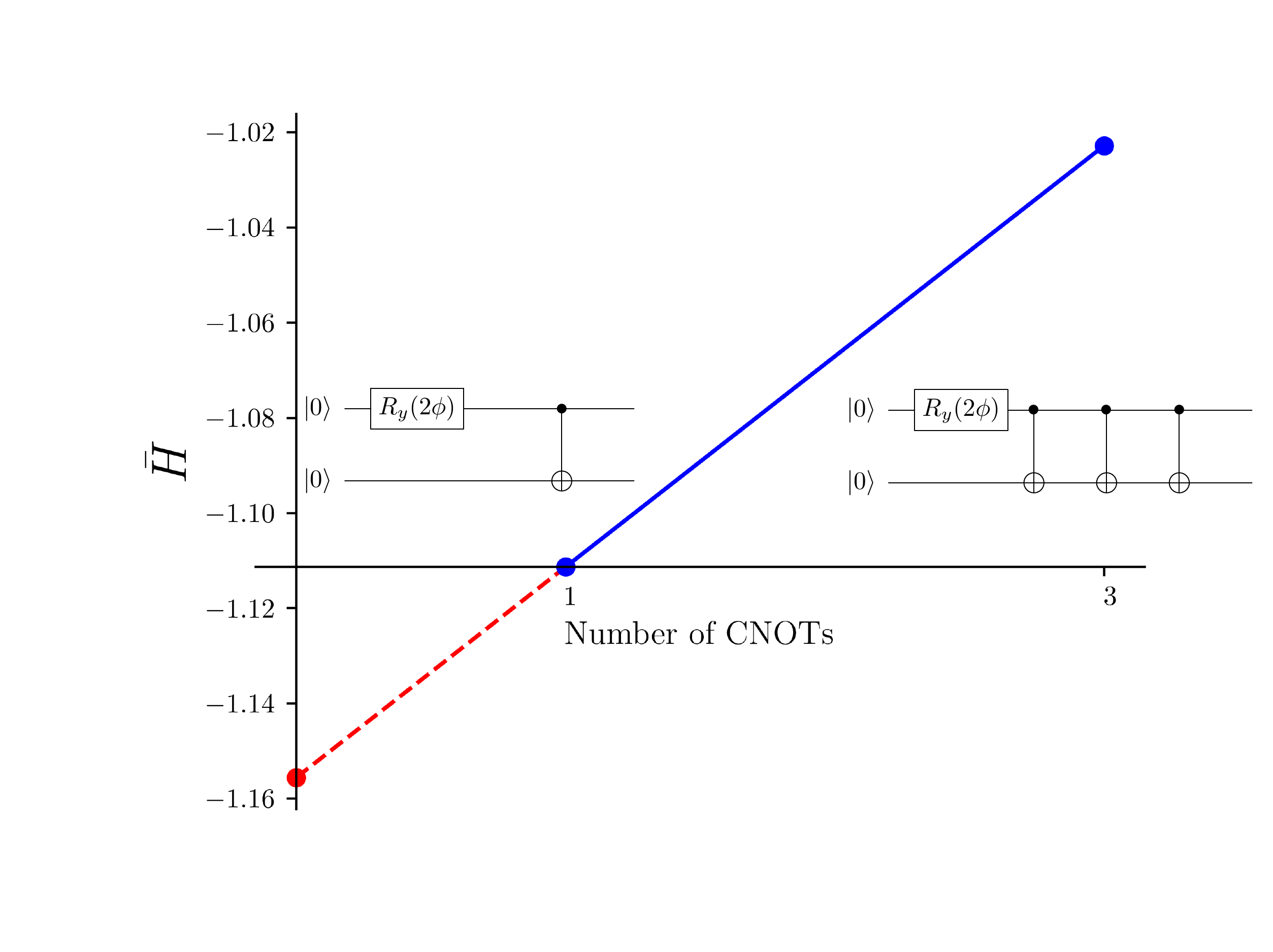}
\caption{Illustration of noise extrapolation method used to mitigate two-qubit errors. Linear extrapolation to zero error from two-qubit gate depolarization for the measured ground state expectation value of the Hamiltonian $H$ with $N=2$, as given in Eq.~\eqref{eq:2qubitHamiltonian}, using measured data when one or three identical {\small CNOT} gates are used to prepare the variational state in Eq.~\eqref{eq:N=2trial}, using the optimum value of the variational parameter $\theta=\pi/8$. In the quantum circuit on the right, each {\small CNOT} in the quantum circuit on the left is replaced by three successive {\small CNOT}s. Because CNOT$^2=1$, the functionality of the two circuits in the absence of noise is the same, but in the presence of noise the circuit with the additional {\small CNOT}s will have larger error.  The error-mitigated estimate for the value of the relevant observable $H$ is obtained by extrapolating the result linearly to zero as a function of the number of {\small CNOT}s. 
} \label{fig:zneScheme} 
\end{figure}

To mitigate two-qubit gate errors, we perform zero noise extrapolation (ZNE) using a linear fit, as discussed in Ref.~\cite{He:2020udd}. In the absence of two-qubit gate errors, inserting two successive identical {\small CNOT} gates anywhere in a circuit does not change the circuit's output. However, in the presence of small {\small CNOT} gate error, inserting additional {\small CNOT}s increases the circuit error by a factor of approximately $r=1+2n$, where $n$ is the number of identity insertions. Figure \ref{fig:zneScheme} shows the circuit that prepares a $N=2$ variational state for the Lipkin model with a single identity insertion. We measure the average value of an observable $\bar{\mathcal{O}}(r)$ for this prepared state as a function of $r$, the number of {\small CNOT}s, and extrapolate linearly to estimate the value of $\mathcal{O}(0)$, the average value of the observable in the absence of {\small CNOT} gate error. 

\begin{figure}
\includegraphics[width=0.49\textwidth]{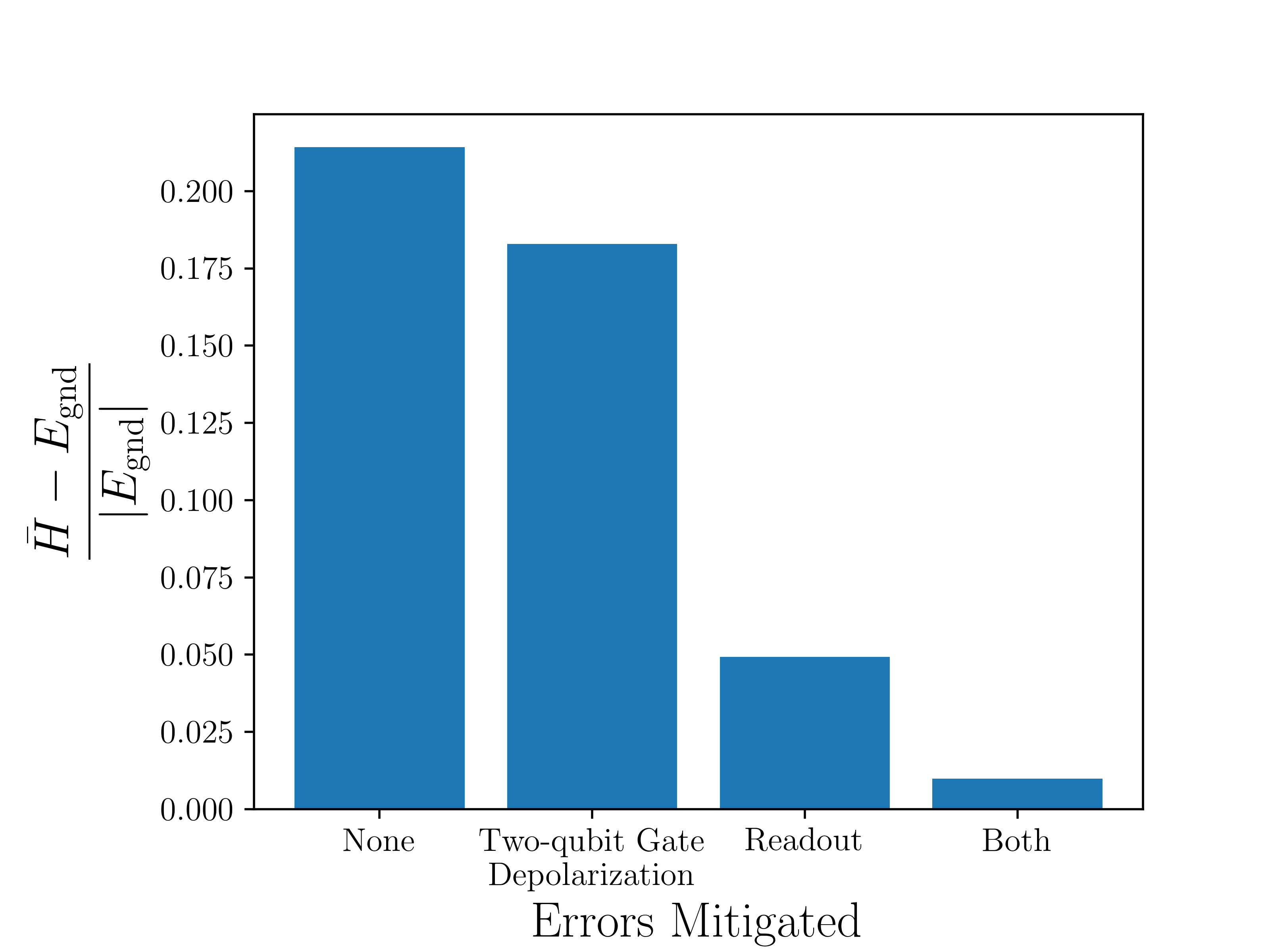}
\caption{
    Performance of error mitigation techniques for estimation of energy for the $N=2$ Lipkin model ground state in Eq.~\eqref{eq:N=2trial}, with fixed coupling $V=1$ and the exact optimum value of the variational parameter $\theta=\pi/8$. All data were calculated on the IBM Q simulator, including all errors listed in Sec.~\ref{sec:errors}.
    The average values of the Hamiltonian operator $H$ as given in Eq.~\eqref{eq:2qubitHamiltonian} for the prepared state are compared using different error mitigation techniques. Deviations from the exact value obtained analytically, $E_\mathrm{gnd}=-\sqrt{2}$, are compared when implementing neither, one, or both of the mitigation techniques for readout and two-qubit gate depolarization errors, as described in Sec.~\ref{subsec:error_mitigation}. 
}
\label{fig:H-vs_miti}
\end{figure}

We characterize the improvements to the accuracy in measured average values of the observables obtained by using error mitigation when the variational parameter $\theta$ is fixed at its known optimum value, $\theta=\pi/8$. 
First, we test how adding each mitigation technique separately to the process of estimating $\braket{H}$ improves the accuracy. We also mitigate both measurement errors and two-qubit gate errors by applying the inverted calibration matrix $M^{-1}$ to the probability distributions obtained from 
each of the circuits in Fig.~\ref{fig:zneScheme}; 
subsequently, results from calculating $\bar{H}$ with each of these circuits can be used to linearly extrapolate a value of $\bar{H}$ with $r=0$, to mitigate two-qubit gate depolarization errors in addition to readout errors. Results comparing these techniques separately and together for $\bar{H}$ are presented in Fig.~\ref{fig:H-vs_miti}, where it can be seen that mitigating both the readout and {\small CNOT} errors improves the accuracy of the results substantially.

\section{Computing observables}
\label{sec:dm}

To be of interest to experiments, such as direct detection of dark matter, we need to not only be able
to compute the energy of an optimal VQE trial for a many-body ground state
with lower complexity and greater accuracy than classical computations,
but also to use
the approximate ground state wave function $\ket{\psi(\theta_{\min})}$ 
to compute physical observables. 
As a stand-in for this goal, we here compute expectation values of observables in the LMG model,
such as $H^n$, $J_1^n$, and $J_0^n$ for $n\in\mathbb{N}$, as given by Eqs.~\eqref{eq:spinH},~\eqref{eq:Jpm}, and~\eqref{eq:Jz}, where $J_1 =(J_++J_-)/2$. In this section, we explore the precision of results obtained from current hardware; specifically, given the availability of readout error mitigation techniques~\cite{QiskitErrorMitigate}, we consider results of simulations with a noise model including all sources of infidelity described in Sec.~\ref{sec:noise} except for readout error, to compare with exact analytic calculations of the LMG model.

\begin{figure}
\centering
\includegraphics[width=0.49\textwidth]{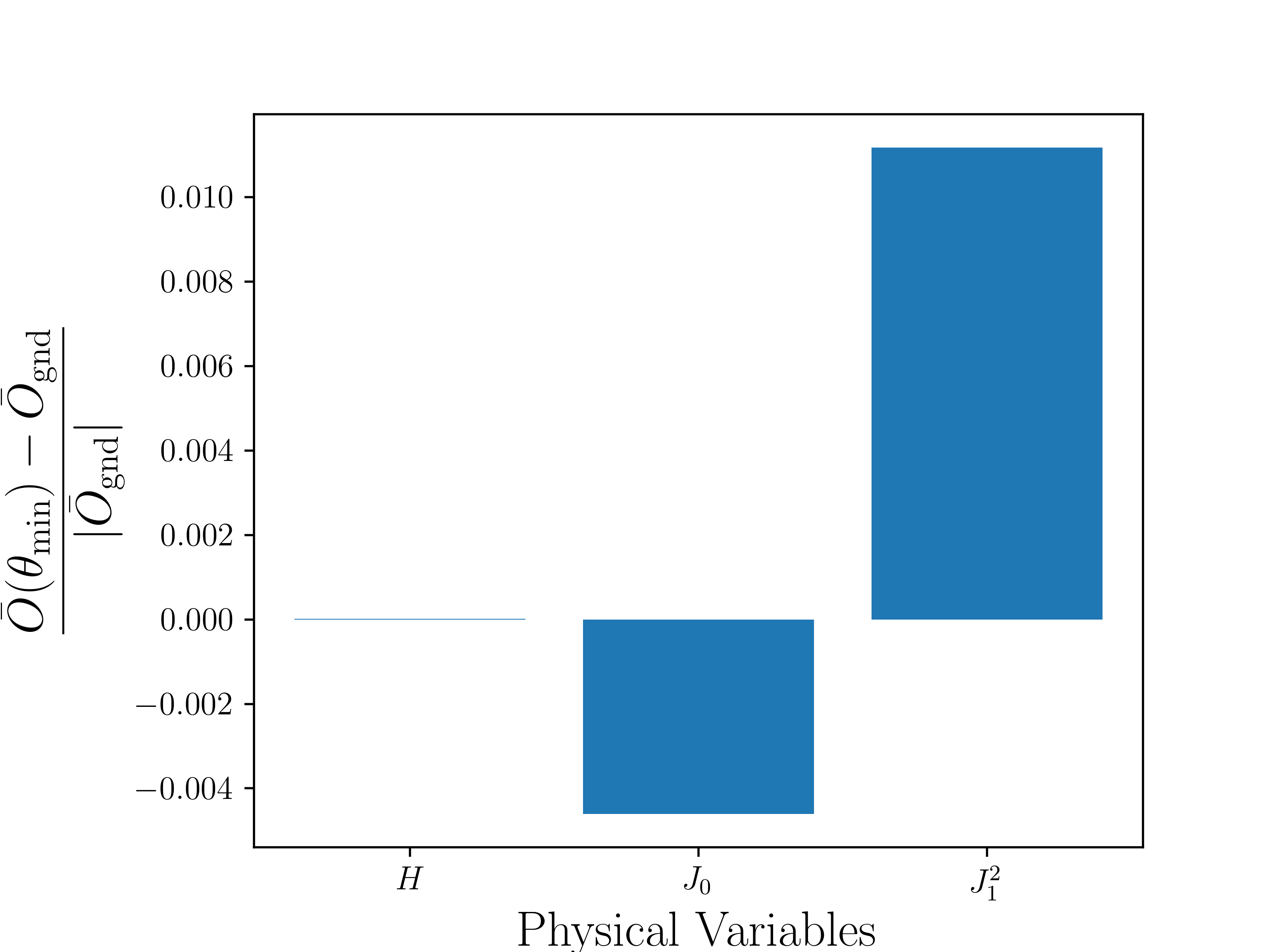} 
\caption{
    Moments of physical quantities used in dark matter detection calculations are calculated using a noise model simulation of the Melbourne processor with no readout error to obtain $\theta_{\min}$ and subsequently using Eq.~\eqref{eq:avgpowerop}, which are compared to exact values obtained analytically. 
    Exact values for these averaged quantities are $\bar{H}_\mathrm{gnd}=-\sqrt{2}$, $(\bar{J_0})_\mathrm{gnd}=-1/\sqrt{2}$, and $(\overline{J_1^2})_\mathrm{gnd}=(2-\sqrt{2})/4$.
} \label{fig:dm}
\end{figure}

As in the preceding calculations, we consider the $N=2$ system with $V=1$. In this case, note that for $O\in\{H,J_0,J_1\}$, $O^3=O$, so we may summarize the behavior of all moments
\begin{equation}
\overline{O^n}(\theta)\equiv\braket{\psi(\theta)|O^n|\psi(\theta)}
\label{eq:avgpowerop}
\end{equation}
for all powers $n\in\mathbb{N}$ using only the values $n=1,2$ with a given state $\psi(\theta)$. Furthermore, by the symmetries $\braket{01|\psi(\theta)}=\braket{10|\psi(\theta)}=0$, as per Eq.~\eqref{eq:N=2trial}, we can observe that $\overline{H^2}(\theta)=\overline{J_0^2}(\theta)=\bar{J_1}(\theta)=0$ exactly for all values of $\theta$. Thus, we need to consider only $\bar{H}(\theta)$, $\bar{J_0}(\theta)$, and $\overline{J_1^2}(\theta)$. The fractional deviation of these averages obtained with the VQE optimal value $\theta=\theta_{\min}$ from results obtained with the exact solution $\theta=\pi/8$ are displayed in Fig.~\ref{fig:dm}.

Results obtained here from VQE calculations may be compared to those obtained classically such as in Ref.~\cite{uqusdb2019}. For example, their classical calculation of the spin-orbit coupling operator averaged over the ground state carries a fractional deviation of $\approx1.8\%$. While the deviation of $\bar{H}(\theta_{\min})$ is significantly smaller in comparison, both $\bar{J_0}(\theta_{\min})$ and $\overline{J_1^2}(\theta_{\min})$ can carry much larger fractional deviations $\sim1\%$. Even with precise calculations of the ground state energy, useful calculations of the averaged values of spin operators such as $J_0$ and $J_1^2$ will require reduction in other noise errors such as 
thermal relaxation of qubits over two-qubit gate operations, as discussed in Sec.~\ref{sec:noise}.
\begin{figure}
\includegraphics[width=0.49\textwidth]{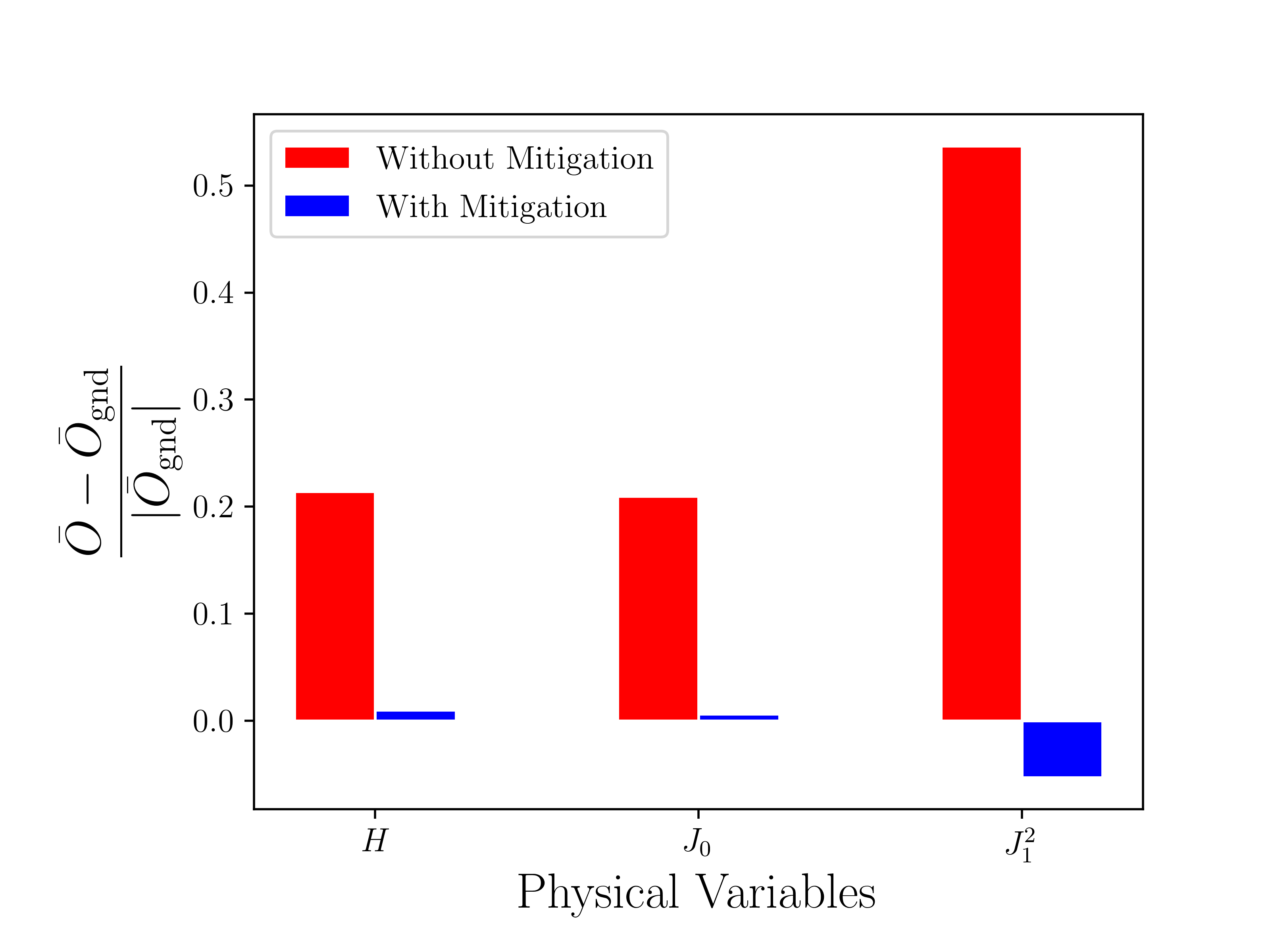}
\caption{
    Performance of error mitigation techniques for estimation of observables for the $N=2$ Lipkin model ground state in Eq.~\eqref{eq:N=2trial} with fixed coupling $V=1$ and the exact optimum value of the variational parameter, $\theta=\pi/8$. All data were calculated using the IBM Q simulator with all errors listed in Sec.~\ref{sec:errors}.
    Moments of physical quantities used in dark matter detection calculations are calculated with and without using mitigation techniques for both two-qubit gate depolarization and readout errors. 
    Exact values for these averaged quantities are $\bar{H}_\mathrm{gnd}=E_\mathrm{gnd}$, $(\bar{J_0})_\mathrm{gnd}=-1/\sqrt{2}$, and $(\overline{J_1^2})_\mathrm{gnd}=(2-\sqrt{2})/4$.
    The error mitigation procedure yields  improvement to the results for the quantities $J_0$ and $J_1^2$ that is similar to that obtained for the energy $H$.
}
\label{fig:J-vs-miti}
\end{figure}

We now characterize the improvements to the accuracy in measured average values of the observables $J_0$ and $J_1^2$ that are obtained by using error mitigation as described in Sec.~\ref{subsec:error_mitigation}.
We
again measure the observables for the variational state with the known optimum value of the variational parameter $\theta=\pi/8$
for circuits in which the {\small CNOT} is replaced by three {\small CNOT}s and extrapolate the results back to obtain the error-mitigated result. 
These results are displayed in Fig.~\ref{fig:J-vs-miti}. 
We find that this mitigation technique yields improvements of an order of magnitude in the accuracy of the measured observables $J_0$ and $J_1^2$. However, significant additional accuracy improvements will still be needed to exceed the $\sim1\%$ precision of existing classical calculations of these observables in more realistic situations, such as results presented in Ref.~\cite{uqusdb2019}.

\section{Summary}
\label{sec:summary}

Appropriate interpretation of the results of experiments, 
including upper bounds, requires reliable models of  target nuclides, including quantified 
uncertainties. Quantum computing has the potential to enable one to go beyond the limitations of classical calculations,  
improving the models as well as understanding the uncertainties in those models. 

Studying the ground state of many-body systems similar to the LMG model can pose a complex quantum problem. 
Highly accurate quantum processors are needed
for quantum computing to yield improvements over classical algorithms.
To identify and assess 
the most significant sources of error in an existing quantum processor,
we develop quantum circuits for VQE calculations on the LMG model and implement the algorithm for the simplest nontrivial case.
We compare VQE results and the exact ground state of the LMG model 
and identify the dominant errors limiting the accuracy of the calculation.
We find that readout error and two-qubit gate errors are the dominant sources of infidelities using current quantum hardware. 
Further, we find that error mitigation techniques improve the accuracy of the calculations substantially. Our results suggest that, given  recent rapid advances in the development of quantum computing hardware~\cite{satzinger2021realizing}, near-term quantum computers could help with the calculation of at least gross properties of nuclear ground states.


$\>$

\section*{Acknowledgments} 

The authors acknowledge support by the U.S. Department of Energy, Office
of Science, Office of High Energy Physics, under Award No.~DE-SC0019465. This work was also supported in part by the National Science Foundation Grant No.~PHY-1806368.
We thank IBM Quantum Experience for making their quantum processors publicly available. 

\bibliography{lipkin_refs}

\begin{thebibliography}{83}%
\makeatletter
\providecommand \@ifxundefined [1]{%
 \@ifx{#1\undefined}
}%
\providecommand \@ifnum [1]{%
 \ifnum #1\expandafter \@firstoftwo
 \else \expandafter \@secondoftwo
 \fi
}%
\providecommand \@ifx [1]{%
 \ifx #1\expandafter \@firstoftwo
 \else \expandafter \@secondoftwo
 \fi
}%
\providecommand \natexlab [1]{#1}%
\providecommand \enquote  [1]{``#1''}%
\providecommand \bibnamefont  [1]{#1}%
\providecommand \bibfnamefont [1]{#1}%
\providecommand \citenamefont [1]{#1}%
\providecommand \href@noop [0]{\@secondoftwo}%
\providecommand \href [0]{\begingroup \@sanitize@url \@href}%
\providecommand \@href[1]{\@@startlink{#1}\@@href}%
\providecommand \@@href[1]{\endgroup#1\@@endlink}%
\providecommand \@sanitize@url [0]{\catcode `\\12\catcode `\$12\catcode
  `\&12\catcode `\#12\catcode `\^12\catcode `\_12\catcode `\%12\relax}%
\providecommand \@@startlink[1]{}%
\providecommand \@@endlink[0]{}%
\providecommand \url  [0]{\begingroup\@sanitize@url \@url }%
\providecommand \@url [1]{\endgroup\@href {#1}{\urlprefix }}%
\providecommand \urlprefix  [0]{URL }%
\providecommand \Eprint [0]{\href }%
\providecommand \doibase [0]{http://dx.doi.org/}%
\providecommand \selectlanguage [0]{\@gobble}%
\providecommand \bibinfo  [0]{\@secondoftwo}%
\providecommand \bibfield  [0]{\@secondoftwo}%
\providecommand \translation [1]{[#1]}%
\providecommand \BibitemOpen [0]{}%
\providecommand \bibitemStop [0]{}%
\providecommand \bibitemNoStop [0]{.\EOS\space}%
\providecommand \EOS [0]{\spacefactor3000\relax}%
\providecommand \BibitemShut  [1]{\csname bibitem#1\endcsname}%
\let\auto@bib@innerbib\@empty
\bibitem [{\citenamefont {Blumenthal}\ \emph {et~al.}(1984)\citenamefont
  {Blumenthal}, \citenamefont {Faber}, \citenamefont {Primack},\ and\
  \citenamefont {Rees}}]{Blumenthal:1984bp}%
  \BibitemOpen
  \bibfield  {author} {\bibinfo {author} {\bibfnamefont {G.~R.}\ \bibnamefont
  {Blumenthal}}, \bibinfo {author} {\bibfnamefont {S.}~\bibnamefont {Faber}},
  \bibinfo {author} {\bibfnamefont {J.~R.}\ \bibnamefont {Primack}}, \ and\
  \bibinfo {author} {\bibfnamefont {M.~J.}\ \bibnamefont {Rees}},\ }\href
  {\doibase 10.1038/311517a0} {\bibfield  {journal} {\bibinfo  {journal}
  {Nature}\ }\textbf {\bibinfo {volume} {311}},\ \bibinfo {pages} {517}
  (\bibinfo {year} {1984})}\BibitemShut {NoStop}%
\bibitem [{\citenamefont {Primack}\ \emph {et~al.}(1988)\citenamefont
  {Primack}, \citenamefont {Seckel},\ and\ \citenamefont
  {Sadoulet}}]{Primack:1988zm}%
  \BibitemOpen
  \bibfield  {author} {\bibinfo {author} {\bibfnamefont {J.~R.}\ \bibnamefont
  {Primack}}, \bibinfo {author} {\bibfnamefont {D.}~\bibnamefont {Seckel}}, \
  and\ \bibinfo {author} {\bibfnamefont {B.}~\bibnamefont {Sadoulet}},\ }\href
  {\doibase 10.1146/annurev.ns.38.120188.003535} {\bibfield  {journal}
  {\bibinfo  {journal} {Annu. Rev. Nucl. Part. Sci.}\ }\textbf {\bibinfo
  {volume} {38}},\ \bibinfo {pages} {751} (\bibinfo {year} {1988})}\BibitemShut
  {NoStop}%
\bibitem [{\citenamefont {Feng}(2010)}]{Feng:2010gw}%
  \BibitemOpen
  \bibfield  {author} {\bibinfo {author} {\bibfnamefont {J.~L.}\ \bibnamefont
  {Feng}},\ }\href {\doibase 10.1146/annurev-astro-082708-101659} {\bibfield
  {journal} {\bibinfo  {journal} {Ann. Rev. Astron. Astrophys.}\ }\textbf
  {\bibinfo {volume} {48}},\ \bibinfo {pages} {495} (\bibinfo {year}
  {2010})}\BibitemShut {NoStop}%
\bibitem [{\citenamefont {Bertone}\ and\ \citenamefont
  {Hooper}(2018)}]{Bertone:2016nfn}%
  \BibitemOpen
  \bibfield  {author} {\bibinfo {author} {\bibfnamefont {G.}~\bibnamefont
  {Bertone}}\ and\ \bibinfo {author} {\bibfnamefont {D.}~\bibnamefont
  {Hooper}},\ }\href {\doibase 10.1103/RevModPhys.90.045002} {\bibfield
  {journal} {\bibinfo  {journal} {Rev. Mod. Phys.}\ }\textbf {\bibinfo {volume}
  {90}},\ \bibinfo {pages} {045002} (\bibinfo {year} {2018})}\BibitemShut
  {NoStop}%
\bibitem [{\citenamefont {Frieman}\ \emph {et~al.}(2008)\citenamefont
  {Frieman}, \citenamefont {Turner},\ and\ \citenamefont
  {Huterer}}]{Frieman:2008sn}%
  \BibitemOpen
  \bibfield  {author} {\bibinfo {author} {\bibfnamefont {J.}~\bibnamefont
  {Frieman}}, \bibinfo {author} {\bibfnamefont {M.}~\bibnamefont {Turner}}, \
  and\ \bibinfo {author} {\bibfnamefont {D.}~\bibnamefont {Huterer}},\ }\href
  {\doibase 10.1146/annurev.astro.46.060407.145243} {\bibfield  {journal}
  {\bibinfo  {journal} {Ann. Rev. Astron. Astrophys.}\ }\textbf {\bibinfo
  {volume} {46}},\ \bibinfo {pages} {385} (\bibinfo {year} {2008})}\BibitemShut
  {NoStop}%
\bibitem [{\citenamefont {Dine}\ and\ \citenamefont
  {Kusenko}(2003)}]{Dine:2003ax}%
  \BibitemOpen
  \bibfield  {author} {\bibinfo {author} {\bibfnamefont {M.}~\bibnamefont
  {Dine}}\ and\ \bibinfo {author} {\bibfnamefont {A.}~\bibnamefont {Kusenko}},\
  }\href {\doibase 10.1103/RevModPhys.76.1} {\bibfield  {journal} {\bibinfo
  {journal} {Rev. Mod. Phys.}\ }\textbf {\bibinfo {volume} {76}},\ \bibinfo
  {pages} {1} (\bibinfo {year} {2003})}\BibitemShut {NoStop}%
\bibitem [{\citenamefont {Kajino}\ \emph {et~al.}(2019)\citenamefont {Kajino},
  \citenamefont {Aoki}, \citenamefont {Balantekin}, \citenamefont {Diehl},
  \citenamefont {Famiano},\ and\ \citenamefont {Mathews}}]{Kajino:2019abv}%
  \BibitemOpen
  \bibfield  {author} {\bibinfo {author} {\bibfnamefont {T.}~\bibnamefont
  {Kajino}}, \bibinfo {author} {\bibfnamefont {W.}~\bibnamefont {Aoki}},
  \bibinfo {author} {\bibfnamefont {A.~B.}\ \bibnamefont {Balantekin}},
  \bibinfo {author} {\bibfnamefont {R.}~\bibnamefont {Diehl}}, \bibinfo
  {author} {\bibfnamefont {M.~A.}\ \bibnamefont {Famiano}}, \ and\ \bibinfo
  {author} {\bibfnamefont {G.~J.}\ \bibnamefont {Mathews}},\ }\href {\doibase
  10.1016/j.ppnp.2019.02.008} {\bibfield  {journal} {\bibinfo  {journal} {Prog.
  Part. Nucl. Phys.}\ }\textbf {\bibinfo {volume} {107}},\ \bibinfo {pages}
  {109} (\bibinfo {year} {2019})}\BibitemShut {NoStop}%
\bibitem [{\citenamefont {Dolinski}\ \emph {et~al.}(2019)\citenamefont
  {Dolinski}, \citenamefont {Poon},\ and\ \citenamefont
  {Rodejohann}}]{Dolinski:2019nrj}%
  \BibitemOpen
  \bibfield  {author} {\bibinfo {author} {\bibfnamefont {M.~J.}\ \bibnamefont
  {Dolinski}}, \bibinfo {author} {\bibfnamefont {A.~W.~P.}\ \bibnamefont
  {Poon}}, \ and\ \bibinfo {author} {\bibfnamefont {W.}~\bibnamefont
  {Rodejohann}},\ }\href {\doibase 10.1146/annurev-nucl-101918-023407}
  {\bibfield  {journal} {\bibinfo  {journal} {Annu. Rev. Nucl. Part. Sci.}\
  }\textbf {\bibinfo {volume} {69}},\ \bibinfo {pages} {219} (\bibinfo {year}
  {2019})}\BibitemShut {NoStop}%
\bibitem [{\citenamefont {Mirizzi}\ \emph {et~al.}(2016)\citenamefont
  {Mirizzi}, \citenamefont {Tamborra}, \citenamefont {Janka}, \citenamefont
  {Saviano}, \citenamefont {Scholberg}, \citenamefont {Bollig}, \citenamefont
  {Hudepohl},\ and\ \citenamefont {Chakraborty}}]{Mirizzi:2015eza}%
  \BibitemOpen
  \bibfield  {author} {\bibinfo {author} {\bibfnamefont {A.}~\bibnamefont
  {Mirizzi}}, \bibinfo {author} {\bibfnamefont {I.}~\bibnamefont {Tamborra}},
  \bibinfo {author} {\bibfnamefont {H.-T.}\ \bibnamefont {Janka}}, \bibinfo
  {author} {\bibfnamefont {N.}~\bibnamefont {Saviano}}, \bibinfo {author}
  {\bibfnamefont {K.}~\bibnamefont {Scholberg}}, \bibinfo {author}
  {\bibfnamefont {R.}~\bibnamefont {Bollig}}, \bibinfo {author} {\bibfnamefont
  {L.}~\bibnamefont {Hudepohl}}, \ and\ \bibinfo {author} {\bibfnamefont
  {S.}~\bibnamefont {Chakraborty}},\ }\href {\doibase
  10.1393/ncr/i2016-10120-8} {\bibfield  {journal} {\bibinfo  {journal} {Riv.
  Nuovo Cim.}\ }\textbf {\bibinfo {volume} {39}},\ \bibinfo {pages} {1}
  (\bibinfo {year} {2016})}\BibitemShut {NoStop}%
\bibitem [{\citenamefont {Schumann}(2019)}]{Schumann:2019eaa}%
  \BibitemOpen
  \bibfield  {author} {\bibinfo {author} {\bibfnamefont {M.}~\bibnamefont
  {Schumann}},\ }\href {\doibase 10.1088/1361-6471/ab2ea5} {\bibfield
  {journal} {\bibinfo  {journal} {J. Phys. G}\ }\textbf {\bibinfo {volume}
  {46}},\ \bibinfo {pages} {103003} (\bibinfo {year} {2019})}\BibitemShut
  {NoStop}%
\bibitem [{\citenamefont {Cannoni}(2013)}]{PhysRevD.87.075014}%
  \BibitemOpen
  \bibfield  {author} {\bibinfo {author} {\bibfnamefont {M.}~\bibnamefont
  {Cannoni}},\ }\href {\doibase 10.1103/PhysRevD.87.075014} {\bibfield
  {journal} {\bibinfo  {journal} {Phys. Rev. D}\ }\textbf {\bibinfo {volume}
  {87}},\ \bibinfo {pages} {075014} (\bibinfo {year} {2013})}\BibitemShut
  {NoStop}%
\bibitem [{\citenamefont {Cerde\~no}\ \emph {et~al.}(2013)\citenamefont
  {Cerde\~no}, \citenamefont {Fornasa}, \citenamefont {Huh},\ and\
  \citenamefont {Peir\'o}}]{PhysRevD.87.023512}%
  \BibitemOpen
  \bibfield  {author} {\bibinfo {author} {\bibfnamefont {D.~G.}\ \bibnamefont
  {Cerde\~no}}, \bibinfo {author} {\bibfnamefont {M.}~\bibnamefont {Fornasa}},
  \bibinfo {author} {\bibfnamefont {J.-H.}\ \bibnamefont {Huh}}, \ and\
  \bibinfo {author} {\bibfnamefont {M.}~\bibnamefont {Peir\'o}},\ }\href
  {\doibase 10.1103/PhysRevD.87.023512} {\bibfield  {journal} {\bibinfo
  {journal} {Phys. Rev. D}\ }\textbf {\bibinfo {volume} {87}},\ \bibinfo
  {pages} {023512} (\bibinfo {year} {2013})}\BibitemShut {NoStop}%
\bibitem [{\citenamefont {Furnstahl}\ \emph {et~al.}(2015)\citenamefont
  {Furnstahl}, \citenamefont {Phillips},\ and\ \citenamefont
  {Wesolowski}}]{furnstahl2015recipe}%
  \BibitemOpen
  \bibfield  {author} {\bibinfo {author} {\bibfnamefont {R.}~\bibnamefont
  {Furnstahl}}, \bibinfo {author} {\bibfnamefont {D.}~\bibnamefont {Phillips}},
  \ and\ \bibinfo {author} {\bibfnamefont {S.}~\bibnamefont {Wesolowski}},\
  }\href@noop {} {\bibfield  {journal} {\bibinfo  {journal} {Journal of Physics
  G: Nuclear and Particle Physics}\ }\textbf {\bibinfo {volume} {42}},\
  \bibinfo {pages} {034028} (\bibinfo {year} {2015})}\BibitemShut {NoStop}%
\bibitem [{\citenamefont {Carlsson}\ \emph {et~al.}(2016)\citenamefont
  {Carlsson}, \citenamefont {Ekstr{\"o}m}, \citenamefont {Forss{\'e}n},
  \citenamefont {Str{\"o}mberg}, \citenamefont {Jansen}, \citenamefont {Lilja},
  \citenamefont {Lindby}, \citenamefont {Mattsson},\ and\ \citenamefont
  {Wendt}}]{carlsson2016uncertainty}%
  \BibitemOpen
  \bibfield  {author} {\bibinfo {author} {\bibfnamefont {B.~D.}\ \bibnamefont
  {Carlsson}}, \bibinfo {author} {\bibfnamefont {A.}~\bibnamefont
  {Ekstr{\"o}m}}, \bibinfo {author} {\bibfnamefont {C.}~\bibnamefont
  {Forss{\'e}n}}, \bibinfo {author} {\bibfnamefont {D.~F.}\ \bibnamefont
  {Str{\"o}mberg}}, \bibinfo {author} {\bibfnamefont {G.~R.}\ \bibnamefont
  {Jansen}}, \bibinfo {author} {\bibfnamefont {O.}~\bibnamefont {Lilja}},
  \bibinfo {author} {\bibfnamefont {M.}~\bibnamefont {Lindby}}, \bibinfo
  {author} {\bibfnamefont {B.~A.}\ \bibnamefont {Mattsson}}, \ and\ \bibinfo
  {author} {\bibfnamefont {K.~A.}\ \bibnamefont {Wendt}},\ }\href@noop {}
  {\bibfield  {journal} {\bibinfo  {journal} {Physical Review X}\ }\textbf
  {\bibinfo {volume} {6}},\ \bibinfo {pages} {011019} (\bibinfo {year}
  {2016})}\BibitemShut {NoStop}%
\bibitem [{\citenamefont {P{\'e}rez}\ \emph {et~al.}(2016)\citenamefont
  {P{\'e}rez}, \citenamefont {Amaro},\ and\ \citenamefont
  {Arriola}}]{perez2016uncertainty}%
  \BibitemOpen
  \bibfield  {author} {\bibinfo {author} {\bibfnamefont {R.~N.}\ \bibnamefont
  {P{\'e}rez}}, \bibinfo {author} {\bibfnamefont {J.}~\bibnamefont {Amaro}}, \
  and\ \bibinfo {author} {\bibfnamefont {E.~R.}\ \bibnamefont {Arriola}},\
  }\href@noop {} {\bibfield  {journal} {\bibinfo  {journal} {International
  Journal of Modern Physics E}\ }\textbf {\bibinfo {volume} {25}},\ \bibinfo
  {pages} {1641009} (\bibinfo {year} {2016})}\BibitemShut {NoStop}%
\bibitem [{\citenamefont {Yoshida}\ \emph {et~al.}(2018)\citenamefont
  {Yoshida}, \citenamefont {Shimizu}, \citenamefont {Togashi},\ and\
  \citenamefont {Otsuka}}]{PhysRevC.98.061301}%
  \BibitemOpen
  \bibfield  {author} {\bibinfo {author} {\bibfnamefont {S.}~\bibnamefont
  {Yoshida}}, \bibinfo {author} {\bibfnamefont {N.}~\bibnamefont {Shimizu}},
  \bibinfo {author} {\bibfnamefont {T.}~\bibnamefont {Togashi}}, \ and\
  \bibinfo {author} {\bibfnamefont {T.}~\bibnamefont {Otsuka}},\ }\href
  {\doibase 10.1103/PhysRevC.98.061301} {\bibfield  {journal} {\bibinfo
  {journal} {Phys. Rev. C}\ }\textbf {\bibinfo {volume} {98}},\ \bibinfo
  {pages} {061301(R)} (\bibinfo {year} {2018})}\BibitemShut {NoStop}%
\bibitem [{\citenamefont {Fox}\ \emph {et~al.}(2020)\citenamefont {Fox},
  \citenamefont {Johnson},\ and\ \citenamefont {Perez}}]{uqusdb2019}%
  \BibitemOpen
  \bibfield  {author} {\bibinfo {author} {\bibfnamefont {J.~M.~R.}\
  \bibnamefont {Fox}}, \bibinfo {author} {\bibfnamefont {C.~W.}\ \bibnamefont
  {Johnson}}, \ and\ \bibinfo {author} {\bibfnamefont {R.~N.}\ \bibnamefont
  {Perez}},\ }\href {\doibase 10.1103/PhysRevC.101.054308} {\bibfield
  {journal} {\bibinfo  {journal} {Phys. Rev. C}\ }\textbf {\bibinfo {volume}
  {101}},\ \bibinfo {pages} {054308} (\bibinfo {year} {2020})}\BibitemShut
  {NoStop}%
\bibitem [{\citenamefont {Hempel}\ \emph {et~al.}(2018)\citenamefont {Hempel},
  \citenamefont {Maier}, \citenamefont {Romero}, \citenamefont {McClean},
  \citenamefont {Monz}, \citenamefont {Shen}, \citenamefont {Jurcevic},
  \citenamefont {Lanyon}, \citenamefont {Love}, \citenamefont {Babbush},
  \citenamefont {Aspuru-Guzik}, \citenamefont {Blatt},\ and\ \citenamefont
  {Roos}}]{Hempel2018}%
  \BibitemOpen
  \bibfield  {author} {\bibinfo {author} {\bibfnamefont {C.}~\bibnamefont
  {Hempel}}, \bibinfo {author} {\bibfnamefont {C.}~\bibnamefont {Maier}},
  \bibinfo {author} {\bibfnamefont {J.}~\bibnamefont {Romero}}, \bibinfo
  {author} {\bibfnamefont {J.}~\bibnamefont {McClean}}, \bibinfo {author}
  {\bibfnamefont {T.}~\bibnamefont {Monz}}, \bibinfo {author} {\bibfnamefont
  {H.}~\bibnamefont {Shen}}, \bibinfo {author} {\bibfnamefont {P.}~\bibnamefont
  {Jurcevic}}, \bibinfo {author} {\bibfnamefont {B.~P.}\ \bibnamefont
  {Lanyon}}, \bibinfo {author} {\bibfnamefont {P.}~\bibnamefont {Love}},
  \bibinfo {author} {\bibfnamefont {R.}~\bibnamefont {Babbush}}, \bibinfo
  {author} {\bibfnamefont {A.}~\bibnamefont {Aspuru-Guzik}}, \bibinfo {author}
  {\bibfnamefont {R.}~\bibnamefont {Blatt}}, \ and\ \bibinfo {author}
  {\bibfnamefont {C.~F.}\ \bibnamefont {Roos}},\ }\href@noop {} {\bibfield
  {journal} {\bibinfo  {journal} {Physical Review X}\ }\textbf {\bibinfo
  {volume} {8}},\ \bibinfo {pages} {031022} (\bibinfo {year}
  {2018})}\BibitemShut {NoStop}%
\bibitem [{\citenamefont {Cao}\ \emph {et~al.}(2019)\citenamefont {Cao},
  \citenamefont {Romero}, \citenamefont {Olson}, \citenamefont {Degroote},
  \citenamefont {Johnson}, \citenamefont {Kieferov{\'a}}, \citenamefont
  {Kivlichan}, \citenamefont {Menke}, \citenamefont {Peropadre}, \citenamefont
  {Sawaya} \emph {et~al.}}]{cao2019quantum}%
  \BibitemOpen
  \bibfield  {author} {\bibinfo {author} {\bibfnamefont {Y.}~\bibnamefont
  {Cao}}, \bibinfo {author} {\bibfnamefont {J.}~\bibnamefont {Romero}},
  \bibinfo {author} {\bibfnamefont {J.~P.}\ \bibnamefont {Olson}}, \bibinfo
  {author} {\bibfnamefont {M.}~\bibnamefont {Degroote}}, \bibinfo {author}
  {\bibfnamefont {P.~D.}\ \bibnamefont {Johnson}}, \bibinfo {author}
  {\bibfnamefont {M.}~\bibnamefont {Kieferov{\'a}}}, \bibinfo {author}
  {\bibfnamefont {I.~D.}\ \bibnamefont {Kivlichan}}, \bibinfo {author}
  {\bibfnamefont {T.}~\bibnamefont {Menke}}, \bibinfo {author} {\bibfnamefont
  {B.}~\bibnamefont {Peropadre}}, \bibinfo {author} {\bibfnamefont {N.~P.}\
  \bibnamefont {Sawaya}},  \emph {et~al.},\ }\href@noop {} {\bibfield
  {journal} {\bibinfo  {journal} {Chemical reviews}\ }\textbf {\bibinfo
  {volume} {119}},\ \bibinfo {pages} {10856} (\bibinfo {year}
  {2019})}\BibitemShut {NoStop}%
\bibitem [{\citenamefont {Arute}\ \emph {et~al.}(2020)\citenamefont {Arute},
  \citenamefont {Arya}, \citenamefont {Babbush}, \citenamefont {Bacon},
  \citenamefont {Bardin}, \citenamefont {Barends}, \citenamefont {Boixo},
  \citenamefont {Broughton}, \citenamefont {Buckley}, \citenamefont {Buell},
  \citenamefont {Burkett}, \citenamefont {Bushnell}, \citenamefont {Chen},
  \citenamefont {Chen}, \citenamefont {Chiaro}, \citenamefont {Collins},
  \citenamefont {Courtney}, \citenamefont {Demura}, \citenamefont {Dunsworth},
  \citenamefont {Farhi} \emph {et~al.}}]{googleHF2020}%
  \BibitemOpen
  \bibfield  {author} {\bibinfo {author} {\bibfnamefont {F.}~\bibnamefont
  {Arute}}, \bibinfo {author} {\bibfnamefont {K.}~\bibnamefont {Arya}},
  \bibinfo {author} {\bibfnamefont {R.}~\bibnamefont {Babbush}}, \bibinfo
  {author} {\bibfnamefont {D.}~\bibnamefont {Bacon}}, \bibinfo {author}
  {\bibfnamefont {J.~C.}\ \bibnamefont {Bardin}}, \bibinfo {author}
  {\bibfnamefont {R.}~\bibnamefont {Barends}}, \bibinfo {author} {\bibfnamefont
  {S.}~\bibnamefont {Boixo}}, \bibinfo {author} {\bibfnamefont
  {M.}~\bibnamefont {Broughton}}, \bibinfo {author} {\bibfnamefont {B.~B.}\
  \bibnamefont {Buckley}}, \bibinfo {author} {\bibfnamefont {D.~A.}\
  \bibnamefont {Buell}}, \bibinfo {author} {\bibfnamefont {B.}~\bibnamefont
  {Burkett}}, \bibinfo {author} {\bibfnamefont {N.}~\bibnamefont {Bushnell}},
  \bibinfo {author} {\bibfnamefont {Y.}~\bibnamefont {Chen}}, \bibinfo {author}
  {\bibfnamefont {Z.}~\bibnamefont {Chen}}, \bibinfo {author} {\bibfnamefont
  {B.}~\bibnamefont {Chiaro}}, \bibinfo {author} {\bibfnamefont
  {R.}~\bibnamefont {Collins}}, \bibinfo {author} {\bibfnamefont
  {W.}~\bibnamefont {Courtney}}, \bibinfo {author} {\bibfnamefont
  {S.}~\bibnamefont {Demura}}, \bibinfo {author} {\bibfnamefont
  {A.}~\bibnamefont {Dunsworth}}, \bibinfo {author} {\bibfnamefont
  {E.}~\bibnamefont {Farhi}},  \emph {et~al.},\ }\href {\doibase
  10.1126/science.abb9811} {\bibfield  {journal} {\bibinfo  {journal}
  {Science}\ }\textbf {\bibinfo {volume} {369}},\ \bibinfo {pages} {1084}
  (\bibinfo {year} {2020})},\ \Eprint
  {http://arxiv.org/abs/\url{https://science.sciencemag.org/content/369/6507/1084.full.pdf}}
  {\url{https://science.sciencemag.org/content/369/6507/1084.full.pdf}}
  \BibitemShut {NoStop}%
\bibitem [{\citenamefont {Dumitrescu}\ \emph {et~al.}(2018)\citenamefont
  {Dumitrescu}, \citenamefont {McCaskey}, \citenamefont {Hagen}, \citenamefont
  {Jansen}, \citenamefont {Morris}, \citenamefont {Papenbrock}, \citenamefont
  {Pooser}, \citenamefont {Dean},\ and\ \citenamefont
  {Lougovski}}]{Dumitrescu:2018njn}%
  \BibitemOpen
  \bibfield  {author} {\bibinfo {author} {\bibfnamefont {E.~F.}\ \bibnamefont
  {Dumitrescu}}, \bibinfo {author} {\bibfnamefont {A.~J.}\ \bibnamefont
  {McCaskey}}, \bibinfo {author} {\bibfnamefont {G.}~\bibnamefont {Hagen}},
  \bibinfo {author} {\bibfnamefont {G.~R.}\ \bibnamefont {Jansen}}, \bibinfo
  {author} {\bibfnamefont {T.~D.}\ \bibnamefont {Morris}}, \bibinfo {author}
  {\bibfnamefont {T.}~\bibnamefont {Papenbrock}}, \bibinfo {author}
  {\bibfnamefont {R.~C.}\ \bibnamefont {Pooser}}, \bibinfo {author}
  {\bibfnamefont {D.~J.}\ \bibnamefont {Dean}}, \ and\ \bibinfo {author}
  {\bibfnamefont {P.}~\bibnamefont {Lougovski}},\ }\href@noop {} {\bibfield
  {journal} {\bibinfo  {journal} {Physical Review Letters}\ }\textbf {\bibinfo
  {volume} {120}},\ \bibinfo {pages} {210501} (\bibinfo {year}
  {2018})}\BibitemShut {NoStop}%
\bibitem [{\citenamefont {Roggero}\ \emph {et~al.}(2020)\citenamefont
  {Roggero}, \citenamefont {Li}, \citenamefont {Carlson}, \citenamefont
  {Gupta},\ and\ \citenamefont {Perdue}}]{PhysRevD.101.074038}%
  \BibitemOpen
  \bibfield  {author} {\bibinfo {author} {\bibfnamefont {A.}~\bibnamefont
  {Roggero}}, \bibinfo {author} {\bibfnamefont {A.~C.~Y.}\ \bibnamefont {Li}},
  \bibinfo {author} {\bibfnamefont {J.}~\bibnamefont {Carlson}}, \bibinfo
  {author} {\bibfnamefont {R.}~\bibnamefont {Gupta}}, \ and\ \bibinfo {author}
  {\bibfnamefont {G.~N.}\ \bibnamefont {Perdue}},\ }\href {\doibase
  10.1103/PhysRevD.101.074038} {\bibfield  {journal} {\bibinfo  {journal}
  {Phys. Rev. D}\ }\textbf {\bibinfo {volume} {101}},\ \bibinfo {pages}
  {074038} (\bibinfo {year} {2020})}\BibitemShut {NoStop}%
\bibitem [{\citenamefont {Kreshchuk}\ \emph {et~al.}(2021)\citenamefont
  {Kreshchuk}, \citenamefont {Jia}, \citenamefont {Kirby}, \citenamefont
  {Goldstein}, \citenamefont {Vary},\ and\ \citenamefont
  {Love}}]{Kreshchuk:2020kcz}%
  \BibitemOpen
  \bibfield  {author} {\bibinfo {author} {\bibfnamefont {M.}~\bibnamefont
  {Kreshchuk}}, \bibinfo {author} {\bibfnamefont {S.}~\bibnamefont {Jia}},
  \bibinfo {author} {\bibfnamefont {W.~M.}\ \bibnamefont {Kirby}}, \bibinfo
  {author} {\bibfnamefont {G.}~\bibnamefont {Goldstein}}, \bibinfo {author}
  {\bibfnamefont {J.~P.}\ \bibnamefont {Vary}}, \ and\ \bibinfo {author}
  {\bibfnamefont {P.~J.}\ \bibnamefont {Love}},\ }\href {\doibase
  10.3390/e23050597} {\bibfield  {journal} {\bibinfo  {journal} {Entropy}\
  }\textbf {\bibinfo {volume} {23}},\ \bibinfo {pages} {597} (\bibinfo {year}
  {2021})},\ \Eprint {http://arxiv.org/abs/2009.07885} {arXiv:2009.07885
  [quant-ph]} \BibitemShut {NoStop}%
\bibitem [{\citenamefont {Mueller}\ \emph {et~al.}(2020)\citenamefont
  {Mueller}, \citenamefont {Tarasov},\ and\ \citenamefont
  {Venugopalan}}]{mueller2020deeply}%
  \BibitemOpen
  \bibfield  {author} {\bibinfo {author} {\bibfnamefont {N.}~\bibnamefont
  {Mueller}}, \bibinfo {author} {\bibfnamefont {A.}~\bibnamefont {Tarasov}}, \
  and\ \bibinfo {author} {\bibfnamefont {R.}~\bibnamefont {Venugopalan}},\
  }\href@noop {} {\bibfield  {journal} {\bibinfo  {journal} {Physical Review
  D}\ }\textbf {\bibinfo {volume} {102}},\ \bibinfo {pages} {016007} (\bibinfo
  {year} {2020})}\BibitemShut {NoStop}%
\bibitem [{\citenamefont {Babbush}\ \emph {et~al.}(2017)\citenamefont
  {Babbush}, \citenamefont {Berry}, \citenamefont {Sanders}, \citenamefont
  {Kivlichan}, \citenamefont {Scherer}, \citenamefont {Wei}, \citenamefont
  {Love},\ and\ \citenamefont {Aspuru-Guzik}}]{babbush2017exponentially}%
  \BibitemOpen
  \bibfield  {author} {\bibinfo {author} {\bibfnamefont {R.}~\bibnamefont
  {Babbush}}, \bibinfo {author} {\bibfnamefont {D.~W.}\ \bibnamefont {Berry}},
  \bibinfo {author} {\bibfnamefont {Y.~R.}\ \bibnamefont {Sanders}}, \bibinfo
  {author} {\bibfnamefont {I.~D.}\ \bibnamefont {Kivlichan}}, \bibinfo {author}
  {\bibfnamefont {A.}~\bibnamefont {Scherer}}, \bibinfo {author} {\bibfnamefont
  {A.~Y.}\ \bibnamefont {Wei}}, \bibinfo {author} {\bibfnamefont {P.~J.}\
  \bibnamefont {Love}}, \ and\ \bibinfo {author} {\bibfnamefont
  {A.}~\bibnamefont {Aspuru-Guzik}},\ }\href@noop {} {\bibfield  {journal}
  {\bibinfo  {journal} {Quantum Science and Technology}\ }\textbf {\bibinfo
  {volume} {3}},\ \bibinfo {pages} {015006} (\bibinfo {year}
  {2017})}\BibitemShut {NoStop}%
\bibitem [{\citenamefont {Ryabinkin}\ \emph {et~al.}(2018)\citenamefont
  {Ryabinkin}, \citenamefont {Yen}, \citenamefont {Genin},\ and\ \citenamefont
  {Izmaylov}}]{ryabinkin2018qubit}%
  \BibitemOpen
  \bibfield  {author} {\bibinfo {author} {\bibfnamefont {I.~G.}\ \bibnamefont
  {Ryabinkin}}, \bibinfo {author} {\bibfnamefont {T.-C.}\ \bibnamefont {Yen}},
  \bibinfo {author} {\bibfnamefont {S.~N.}\ \bibnamefont {Genin}}, \ and\
  \bibinfo {author} {\bibfnamefont {A.~F.}\ \bibnamefont {Izmaylov}},\
  }\href@noop {} {\bibfield  {journal} {\bibinfo  {journal} {Journal of
  chemical theory and computation}\ }\textbf {\bibinfo {volume} {14}},\
  \bibinfo {pages} {6317} (\bibinfo {year} {2018})}\BibitemShut {NoStop}%
\bibitem [{\citenamefont {Romero}\ \emph {et~al.}(2018)\citenamefont {Romero},
  \citenamefont {Babbush}, \citenamefont {McClean}, \citenamefont {Hempel},
  \citenamefont {Love},\ and\ \citenamefont
  {Aspuru-Guzik}}]{romero2018strategies}%
  \BibitemOpen
  \bibfield  {author} {\bibinfo {author} {\bibfnamefont {J.}~\bibnamefont
  {Romero}}, \bibinfo {author} {\bibfnamefont {R.}~\bibnamefont {Babbush}},
  \bibinfo {author} {\bibfnamefont {J.~R.}\ \bibnamefont {McClean}}, \bibinfo
  {author} {\bibfnamefont {C.}~\bibnamefont {Hempel}}, \bibinfo {author}
  {\bibfnamefont {P.~J.}\ \bibnamefont {Love}}, \ and\ \bibinfo {author}
  {\bibfnamefont {A.}~\bibnamefont {Aspuru-Guzik}},\ }\href@noop {} {\bibfield
  {journal} {\bibinfo  {journal} {Quantum Science and Technology}\ }\textbf
  {\bibinfo {volume} {4}},\ \bibinfo {pages} {014008} (\bibinfo {year}
  {2018})}\BibitemShut {NoStop}%
\bibitem [{\citenamefont {Lipkin}\ \emph {et~al.}(1965)\citenamefont {Lipkin},
  \citenamefont {Meshkov},\ and\ \citenamefont {Glick}}]{Lipkin:1964yk}%
  \BibitemOpen
  \bibfield  {author} {\bibinfo {author} {\bibfnamefont {H.~J.}\ \bibnamefont
  {Lipkin}}, \bibinfo {author} {\bibfnamefont {N.}~\bibnamefont {Meshkov}}, \
  and\ \bibinfo {author} {\bibfnamefont {A.~J.}\ \bibnamefont {Glick}},\ }\href
  {\doibase 10.1016/0029-5582(65)90862-X} {\bibfield  {journal} {\bibinfo
  {journal} {Nucl. Phys.}\ }\textbf {\bibinfo {volume} {62}},\ \bibinfo {pages}
  {188} (\bibinfo {year} {1965})}\BibitemShut {NoStop}%
\bibitem [{\citenamefont {Peruzzo}\ \emph {et~al.}(2014)\citenamefont
  {Peruzzo}, \citenamefont {McClean}, \citenamefont {Shadbolt}, \citenamefont
  {Yung}, \citenamefont {Zhou}, \citenamefont {Love}, \citenamefont
  {Aspuru-Guzik},\ and\ \citenamefont {O'Brien}}]{peruzzo2014variational}%
  \BibitemOpen
  \bibfield  {author} {\bibinfo {author} {\bibfnamefont {A.}~\bibnamefont
  {Peruzzo}}, \bibinfo {author} {\bibfnamefont {J.}~\bibnamefont {McClean}},
  \bibinfo {author} {\bibfnamefont {P.}~\bibnamefont {Shadbolt}}, \bibinfo
  {author} {\bibfnamefont {M.-H.}\ \bibnamefont {Yung}}, \bibinfo {author}
  {\bibfnamefont {X.-Q.}\ \bibnamefont {Zhou}}, \bibinfo {author}
  {\bibfnamefont {P.}~\bibnamefont {Love}}, \bibinfo {author} {\bibfnamefont
  {A.}~\bibnamefont {Aspuru-Guzik}}, \ and\ \bibinfo {author} {\bibfnamefont
  {J.~L.}\ \bibnamefont {O'Brien}},\ }\href@noop {} {\bibfield  {journal}
  {\bibinfo  {journal} {Nature Communications}\ }\textbf {\bibinfo {volume}
  {5}},\ \bibinfo {pages} {4213} (\bibinfo {year} {2014})}\BibitemShut
  {NoStop}%
\bibitem [{\citenamefont {Temme}\ \emph {et~al.}(2017)\citenamefont {Temme},
  \citenamefont {Bravyi},\ and\ \citenamefont {Gambetta}}]{temme2017error}%
  \BibitemOpen
  \bibfield  {author} {\bibinfo {author} {\bibfnamefont {K.}~\bibnamefont
  {Temme}}, \bibinfo {author} {\bibfnamefont {S.}~\bibnamefont {Bravyi}}, \
  and\ \bibinfo {author} {\bibfnamefont {J.~M.}\ \bibnamefont {Gambetta}},\
  }\href@noop {} {\bibfield  {journal} {\bibinfo  {journal} {Physical Review
  Letters}\ }\textbf {\bibinfo {volume} {119}},\ \bibinfo {pages} {180509}
  (\bibinfo {year} {2017})}\BibitemShut {NoStop}%
\bibitem [{\citenamefont {He}\ \emph {et~al.}(2020)\citenamefont {He},
  \citenamefont {Nachman}, \citenamefont {de~Jong},\ and\ \citenamefont
  {Bauer}}]{He:2020udd}%
  \BibitemOpen
  \bibfield  {author} {\bibinfo {author} {\bibfnamefont {A.}~\bibnamefont
  {He}}, \bibinfo {author} {\bibfnamefont {B.}~\bibnamefont {Nachman}},
  \bibinfo {author} {\bibfnamefont {W.~A.}\ \bibnamefont {de~Jong}}, \ and\
  \bibinfo {author} {\bibfnamefont {C.~W.}\ \bibnamefont {Bauer}},\ }\href
  {\doibase 10.1103/PhysRevA.102.012426} {\bibfield  {journal} {\bibinfo
  {journal} {Phys. Rev. A}\ }\textbf {\bibinfo {volume} {102}},\ \bibinfo
  {pages} {012426} (\bibinfo {year} {2020})}\BibitemShut {NoStop}%
\bibitem [{\citenamefont {Bertone}\ \emph {et~al.}(2005)\citenamefont
  {Bertone}, \citenamefont {Hooper},\ and\ \citenamefont
  {Silk}}]{Bertone:2004pz}%
  \BibitemOpen
  \bibfield  {author} {\bibinfo {author} {\bibfnamefont {G.}~\bibnamefont
  {Bertone}}, \bibinfo {author} {\bibfnamefont {D.}~\bibnamefont {Hooper}}, \
  and\ \bibinfo {author} {\bibfnamefont {J.}~\bibnamefont {Silk}},\ }\href@noop
  {} {\bibfield  {journal} {\bibinfo  {journal} {Physics Reports}\ }\textbf
  {\bibinfo {volume} {405}},\ \bibinfo {pages} {279} (\bibinfo {year}
  {2005})}\BibitemShut {NoStop}%
\bibitem [{\citenamefont {Jungman}\ \emph {et~al.}(1996)\citenamefont
  {Jungman}, \citenamefont {Kamionkowski},\ and\ \citenamefont
  {Griest}}]{Jungman:1995df}%
  \BibitemOpen
  \bibfield  {author} {\bibinfo {author} {\bibfnamefont {G.}~\bibnamefont
  {Jungman}}, \bibinfo {author} {\bibfnamefont {M.}~\bibnamefont
  {Kamionkowski}}, \ and\ \bibinfo {author} {\bibfnamefont {K.}~\bibnamefont
  {Griest}},\ }\href {\doibase 10.1016/0370-1573(95)00058-5} {\bibfield
  {journal} {\bibinfo  {journal} {Phys. Rep.}\ }\textbf {\bibinfo {volume}
  {267}},\ \bibinfo {pages} {195} (\bibinfo {year} {1996})}\BibitemShut
  {NoStop}%
\bibitem [{\citenamefont {Sadoulet}(1999)}]{RevModPhys.71.S197}%
  \BibitemOpen
  \bibfield  {author} {\bibinfo {author} {\bibfnamefont {B.}~\bibnamefont
  {Sadoulet}},\ }\href {\doibase 10.1103/RevModPhys.71.S197} {\bibfield
  {journal} {\bibinfo  {journal} {Rev. Mod. Phys.}\ }\textbf {\bibinfo {volume}
  {71}},\ \bibinfo {pages} {S197} (\bibinfo {year} {1999})}\BibitemShut
  {NoStop}%
\bibitem [{\citenamefont {Goodman}\ and\ \citenamefont
  {Witten}(1985)}]{PhysRevD.31.3059}%
  \BibitemOpen
  \bibfield  {author} {\bibinfo {author} {\bibfnamefont {M.~W.}\ \bibnamefont
  {Goodman}}\ and\ \bibinfo {author} {\bibfnamefont {E.}~\bibnamefont
  {Witten}},\ }\href {\doibase 10.1103/PhysRevD.31.3059} {\bibfield  {journal}
  {\bibinfo  {journal} {Phys. Rev. D}\ }\textbf {\bibinfo {volume} {31}},\
  \bibinfo {pages} {3059} (\bibinfo {year} {1985})}\BibitemShut {NoStop}%
\bibitem [{\citenamefont {Fitzpatrick}\ \emph {et~al.}(2013)\citenamefont
  {Fitzpatrick}, \citenamefont {Haxton}, \citenamefont {Katz}, \citenamefont
  {Lubbers},\ and\ \citenamefont {Xu}}]{Fitzpatrick:2012ix}%
  \BibitemOpen
  \bibfield  {author} {\bibinfo {author} {\bibfnamefont {A.}~\bibnamefont
  {Fitzpatrick}}, \bibinfo {author} {\bibfnamefont {W.}~\bibnamefont {Haxton}},
  \bibinfo {author} {\bibfnamefont {E.}~\bibnamefont {Katz}}, \bibinfo {author}
  {\bibfnamefont {N.}~\bibnamefont {Lubbers}}, \ and\ \bibinfo {author}
  {\bibfnamefont {Y.}~\bibnamefont {Xu}},\ }\href {\doibase
  10.1088/1475-7516/2013/02/004} {\bibfield  {journal} {\bibinfo  {journal}
  {JCAP}\ }\textbf {\bibinfo {volume} {02}},\ \bibinfo {pages} {004} (\bibinfo
  {year} {2013})}\BibitemShut {NoStop}%
\bibitem [{\citenamefont {Anand}\ \emph {et~al.}(2014)\citenamefont {Anand},
  \citenamefont {Fitzpatrick},\ and\ \citenamefont {Haxton}}]{Anand:2013yka}%
  \BibitemOpen
  \bibfield  {author} {\bibinfo {author} {\bibfnamefont {N.}~\bibnamefont
  {Anand}}, \bibinfo {author} {\bibfnamefont {A.~L.}\ \bibnamefont
  {Fitzpatrick}}, \ and\ \bibinfo {author} {\bibfnamefont {W.}~\bibnamefont
  {Haxton}},\ }\href@noop {} {\bibfield  {journal} {\bibinfo  {journal}
  {Physical Review C}\ }\textbf {\bibinfo {volume} {89}},\ \bibinfo {pages}
  {065501} (\bibinfo {year} {2014})}\BibitemShut {NoStop}%
\bibitem [{\citenamefont {Vietze}\ \emph {et~al.}(2015)\citenamefont {Vietze},
  \citenamefont {Klos}, \citenamefont {Men\'endez}, \citenamefont {Haxton},\
  and\ \citenamefont {Schwenk}}]{Vietze:2014vsa}%
  \BibitemOpen
  \bibfield  {author} {\bibinfo {author} {\bibfnamefont {L.}~\bibnamefont
  {Vietze}}, \bibinfo {author} {\bibfnamefont {P.}~\bibnamefont {Klos}},
  \bibinfo {author} {\bibfnamefont {J.}~\bibnamefont {Men\'endez}}, \bibinfo
  {author} {\bibfnamefont {W.}~\bibnamefont {Haxton}}, \ and\ \bibinfo {author}
  {\bibfnamefont {A.}~\bibnamefont {Schwenk}},\ }\href {\doibase
  10.1103/PhysRevD.91.043520} {\bibfield  {journal} {\bibinfo  {journal} {Phys.
  Rev. D}\ }\textbf {\bibinfo {volume} {91}},\ \bibinfo {pages} {043520}
  (\bibinfo {year} {2015})}\BibitemShut {NoStop}%
\bibitem [{\citenamefont {Fieguth}\ \emph {et~al.}(2018)\citenamefont
  {Fieguth}, \citenamefont {Hoferichter}, \citenamefont {Klos}, \citenamefont
  {Men\'endez}, \citenamefont {Schwenk},\ and\ \citenamefont
  {Weinheimer}}]{Fieguth:2018vob}%
  \BibitemOpen
  \bibfield  {author} {\bibinfo {author} {\bibfnamefont {A.}~\bibnamefont
  {Fieguth}}, \bibinfo {author} {\bibfnamefont {M.}~\bibnamefont
  {Hoferichter}}, \bibinfo {author} {\bibfnamefont {P.}~\bibnamefont {Klos}},
  \bibinfo {author} {\bibfnamefont {J.}~\bibnamefont {Men\'endez}}, \bibinfo
  {author} {\bibfnamefont {A.}~\bibnamefont {Schwenk}}, \ and\ \bibinfo
  {author} {\bibfnamefont {C.}~\bibnamefont {Weinheimer}},\ }\href {\doibase
  10.1103/PhysRevD.97.103532} {\bibfield  {journal} {\bibinfo  {journal} {Phys.
  Rev. D}\ }\textbf {\bibinfo {volume} {97}},\ \bibinfo {pages} {103532}
  (\bibinfo {year} {2018})}\BibitemShut {NoStop}%
\bibitem [{\citenamefont {Hoferichter}\ \emph {et~al.}(2020)\citenamefont
  {Hoferichter}, \citenamefont {Men\'endez},\ and\ \citenamefont
  {Schwenk}}]{Hoferichter:2020osn}%
  \BibitemOpen
  \bibfield  {author} {\bibinfo {author} {\bibfnamefont {M.}~\bibnamefont
  {Hoferichter}}, \bibinfo {author} {\bibfnamefont {J.}~\bibnamefont
  {Men\'endez}}, \ and\ \bibinfo {author} {\bibfnamefont {A.}~\bibnamefont
  {Schwenk}},\ }\href@noop {} {\bibfield  {journal} {\bibinfo  {journal} {Phys.
  Rev. D}\ }\textbf {\bibinfo {volume} {102}},\ \bibinfo {pages} {074018}
  (\bibinfo {year} {2020})}\BibitemShut {NoStop}%
\bibitem [{\citenamefont {Alsum}(2020)}]{alsum2020effective}%
  \BibitemOpen
  \bibfield  {author} {\bibinfo {author} {\bibfnamefont {S.~K.}\ \bibnamefont
  {Alsum}},\ }\emph {\bibinfo {title} {Effective Field Theory Search Results
  from the LUX Run 4 Data Set, and Construction of the LZ System Test
  Platforms}},\ \href@noop {} {Ph.D. thesis},\ \bibinfo  {school} {The
  University of Wisconsin-Madison} (\bibinfo {year} {2020})\BibitemShut
  {NoStop}%
\bibitem [{\citenamefont {Navr{\'a}til}\ \emph {et~al.}(2000)\citenamefont
  {Navr{\'a}til}, \citenamefont {Vary},\ and\ \citenamefont
  {Barrett}}]{navratil2000large}%
  \BibitemOpen
  \bibfield  {author} {\bibinfo {author} {\bibfnamefont {P.}~\bibnamefont
  {Navr{\'a}til}}, \bibinfo {author} {\bibfnamefont {J.}~\bibnamefont {Vary}},
  \ and\ \bibinfo {author} {\bibfnamefont {B.}~\bibnamefont {Barrett}},\
  }\href@noop {} {\bibfield  {journal} {\bibinfo  {journal} {Physical Review
  C}\ }\textbf {\bibinfo {volume} {62}},\ \bibinfo {pages} {054311} (\bibinfo
  {year} {2000})}\BibitemShut {NoStop}%
\bibitem [{\citenamefont {Barrett}\ \emph {et~al.}(2013)\citenamefont
  {Barrett}, \citenamefont {Navr{\'a}til},\ and\ \citenamefont
  {Vary}}]{barrett2013ab}%
  \BibitemOpen
  \bibfield  {author} {\bibinfo {author} {\bibfnamefont {B.~R.}\ \bibnamefont
  {Barrett}}, \bibinfo {author} {\bibfnamefont {P.}~\bibnamefont
  {Navr{\'a}til}}, \ and\ \bibinfo {author} {\bibfnamefont {J.~P.}\
  \bibnamefont {Vary}},\ }\href@noop {} {\bibfield  {journal} {\bibinfo
  {journal} {Progress in Particle and Nuclear Physics}\ }\textbf {\bibinfo
  {volume} {69}},\ \bibinfo {pages} {131} (\bibinfo {year} {2013})}\BibitemShut
  {NoStop}%
\bibitem [{\citenamefont {Hagen}\ \emph {et~al.}(2010)\citenamefont {Hagen},
  \citenamefont {Papenbrock}, \citenamefont {Dean},\ and\ \citenamefont
  {Hjorth-Jensen}}]{hagen2010ab}%
  \BibitemOpen
  \bibfield  {author} {\bibinfo {author} {\bibfnamefont {G.}~\bibnamefont
  {Hagen}}, \bibinfo {author} {\bibfnamefont {T.}~\bibnamefont {Papenbrock}},
  \bibinfo {author} {\bibfnamefont {D.~J.}\ \bibnamefont {Dean}}, \ and\
  \bibinfo {author} {\bibfnamefont {M.}~\bibnamefont {Hjorth-Jensen}},\
  }\href@noop {} {\bibfield  {journal} {\bibinfo  {journal} {Physical Review
  C}\ }\textbf {\bibinfo {volume} {82}},\ \bibinfo {pages} {034330} (\bibinfo
  {year} {2010})}\BibitemShut {NoStop}%
\bibitem [{\citenamefont {Brussard}\ and\ \citenamefont
  {Glaudemans}(1977)}]{BG77}%
  \BibitemOpen
  \bibfield  {author} {\bibinfo {author} {\bibfnamefont {P.~J.}\ \bibnamefont
  {Brussard}}\ and\ \bibinfo {author} {\bibfnamefont {P.~W.~M.}\ \bibnamefont
  {Glaudemans}},\ }\href@noop {} {\emph {\bibinfo {title} {Shell-Model
  Applications in Nuclear Spectroscopy}}}\ (\bibinfo  {publisher}
  {North-Holland Publishing Company, Amsterdam},\ \bibinfo {year}
  {1977})\BibitemShut {NoStop}%
\bibitem [{\citenamefont {Brown}\ and\ \citenamefont
  {Wildenthal}(1988)}]{br88}%
  \BibitemOpen
  \bibfield  {author} {\bibinfo {author} {\bibfnamefont {B.~A.}\ \bibnamefont
  {Brown}}\ and\ \bibinfo {author} {\bibfnamefont {B.~H.}\ \bibnamefont
  {Wildenthal}},\ }\href@noop {} {\bibfield  {journal} {\bibinfo  {journal}
  {{Annu. Rev. Nuc. Part. Sci.}}\ }\textbf {\bibinfo {volume} {38}},\ \bibinfo
  {pages} {29} (\bibinfo {year} {1988})}\BibitemShut {NoStop}%
\bibitem [{\citenamefont {Caurier}\ \emph {et~al.}(2005)\citenamefont
  {Caurier}, \citenamefont {Martinez-Pinedo}, \citenamefont {Nowacki},
  \citenamefont {Poves},\ and\ \citenamefont {Zuker}}]{ca05}%
  \BibitemOpen
  \bibfield  {author} {\bibinfo {author} {\bibfnamefont {E.}~\bibnamefont
  {Caurier}}, \bibinfo {author} {\bibfnamefont {G.}~\bibnamefont
  {Martinez-Pinedo}}, \bibinfo {author} {\bibfnamefont {F.}~\bibnamefont
  {Nowacki}}, \bibinfo {author} {\bibfnamefont {A.}~\bibnamefont {Poves}}, \
  and\ \bibinfo {author} {\bibfnamefont {A.~P.}\ \bibnamefont {Zuker}},\
  }\href@noop {} {\bibfield  {journal} {\bibinfo  {journal} {{Rev. Mod.
  Phys.}}\ }\textbf {\bibinfo {volume} {77}},\ \bibinfo {pages} {427} (\bibinfo
  {year} {2005})}\BibitemShut {NoStop}%
\bibitem [{\citenamefont {Stroberg}\ \emph {et~al.}(2019)\citenamefont
  {Stroberg}, \citenamefont {Hergert}, \citenamefont {Bogner},\ and\
  \citenamefont {Holt}}]{doi:10.1146/annurev-nucl-101917-021120}%
  \BibitemOpen
  \bibfield  {author} {\bibinfo {author} {\bibfnamefont {S.~R.}\ \bibnamefont
  {Stroberg}}, \bibinfo {author} {\bibfnamefont {H.}~\bibnamefont {Hergert}},
  \bibinfo {author} {\bibfnamefont {S.~K.}\ \bibnamefont {Bogner}}, \ and\
  \bibinfo {author} {\bibfnamefont {J.~D.}\ \bibnamefont {Holt}},\ }\href
  {\doibase 10.1146/annurev-nucl-101917-021120} {\bibfield  {journal} {\bibinfo
   {journal} {{Annu. Rev. Nuc. Part. Sci.}}\ }\textbf {\bibinfo {volume}
  {69}},\ \bibinfo {pages} {307} (\bibinfo {year} {2019})}\BibitemShut
  {NoStop}%
\bibitem [{\citenamefont {Pacheco}\ and\ \citenamefont
  {Strottman}(1989)}]{PhysRevD.40.2131}%
  \BibitemOpen
  \bibfield  {author} {\bibinfo {author} {\bibfnamefont {A.~F.}\ \bibnamefont
  {Pacheco}}\ and\ \bibinfo {author} {\bibfnamefont {D.}~\bibnamefont
  {Strottman}},\ }\href {\doibase 10.1103/PhysRevD.40.2131} {\bibfield
  {journal} {\bibinfo  {journal} {Phys. Rev. D}\ }\textbf {\bibinfo {volume}
  {40}},\ \bibinfo {pages} {2131} (\bibinfo {year} {1989})}\BibitemShut
  {NoStop}%
\bibitem [{\citenamefont {Ressell}\ \emph {et~al.}(1993)\citenamefont
  {Ressell}, \citenamefont {Aufderheide}, \citenamefont {Bloom}, \citenamefont
  {Griest}, \citenamefont {Mathews},\ and\ \citenamefont
  {Resler}}]{PhysRevD.48.5519}%
  \BibitemOpen
  \bibfield  {author} {\bibinfo {author} {\bibfnamefont {M.~T.}\ \bibnamefont
  {Ressell}}, \bibinfo {author} {\bibfnamefont {M.~B.}\ \bibnamefont
  {Aufderheide}}, \bibinfo {author} {\bibfnamefont {S.~D.}\ \bibnamefont
  {Bloom}}, \bibinfo {author} {\bibfnamefont {K.}~\bibnamefont {Griest}},
  \bibinfo {author} {\bibfnamefont {G.~J.}\ \bibnamefont {Mathews}}, \ and\
  \bibinfo {author} {\bibfnamefont {D.~A.}\ \bibnamefont {Resler}},\ }\href
  {\doibase 10.1103/PhysRevD.48.5519} {\bibfield  {journal} {\bibinfo
  {journal} {Phys. Rev. D}\ }\textbf {\bibinfo {volume} {48}},\ \bibinfo
  {pages} {5519} (\bibinfo {year} {1993})}\BibitemShut {NoStop}%
\bibitem [{\citenamefont {Pirinen}\ \emph {et~al.}(2016)\citenamefont
  {Pirinen}, \citenamefont {Srivastava}, \citenamefont {Suhonen},\ and\
  \citenamefont {Kortelainen}}]{PhysRevD.93.095012}%
  \BibitemOpen
  \bibfield  {author} {\bibinfo {author} {\bibfnamefont {P.}~\bibnamefont
  {Pirinen}}, \bibinfo {author} {\bibfnamefont {P.~C.}\ \bibnamefont
  {Srivastava}}, \bibinfo {author} {\bibfnamefont {J.}~\bibnamefont {Suhonen}},
  \ and\ \bibinfo {author} {\bibfnamefont {M.}~\bibnamefont {Kortelainen}},\
  }\href {\doibase 10.1103/PhysRevD.93.095012} {\bibfield  {journal} {\bibinfo
  {journal} {Phys. Rev. D}\ }\textbf {\bibinfo {volume} {93}},\ \bibinfo
  {pages} {095012} (\bibinfo {year} {2016})}\BibitemShut {NoStop}%
\bibitem [{\citenamefont {Men\'endez}\ \emph {et~al.}(2012)\citenamefont
  {Men\'endez}, \citenamefont {Gazit},\ and\ \citenamefont
  {Schwenk}}]{PhysRevD.86.103511}%
  \BibitemOpen
  \bibfield  {author} {\bibinfo {author} {\bibfnamefont {J.}~\bibnamefont
  {Men\'endez}}, \bibinfo {author} {\bibfnamefont {D.}~\bibnamefont {Gazit}}, \
  and\ \bibinfo {author} {\bibfnamefont {A.}~\bibnamefont {Schwenk}},\ }\href
  {\doibase 10.1103/PhysRevD.86.103511} {\bibfield  {journal} {\bibinfo
  {journal} {Phys. Rev. D}\ }\textbf {\bibinfo {volume} {86}},\ \bibinfo
  {pages} {103511} (\bibinfo {year} {2012})}\BibitemShut {NoStop}%
\bibitem [{\citenamefont {Klos}\ \emph {et~al.}(2013)\citenamefont {Klos},
  \citenamefont {Men\'endez}, \citenamefont {Gazit},\ and\ \citenamefont
  {Schwenk}}]{PhysRevD.88.083516}%
  \BibitemOpen
  \bibfield  {author} {\bibinfo {author} {\bibfnamefont {P.}~\bibnamefont
  {Klos}}, \bibinfo {author} {\bibfnamefont {J.}~\bibnamefont {Men\'endez}},
  \bibinfo {author} {\bibfnamefont {D.}~\bibnamefont {Gazit}}, \ and\ \bibinfo
  {author} {\bibfnamefont {A.}~\bibnamefont {Schwenk}},\ }\href {\doibase
  10.1103/PhysRevD.88.083516} {\bibfield  {journal} {\bibinfo  {journal} {Phys.
  Rev. D}\ }\textbf {\bibinfo {volume} {88}},\ \bibinfo {pages} {083516}
  (\bibinfo {year} {2013})}\BibitemShut {NoStop}%
\bibitem [{\citenamefont {Baudis}\ \emph {et~al.}(2013)\citenamefont {Baudis},
  \citenamefont {Kessler}, \citenamefont {Klos}, \citenamefont {Lang},
  \citenamefont {Men\'endez}, \citenamefont {Reichard},\ and\ \citenamefont
  {Schwenk}}]{PhysRevD.88.115014}%
  \BibitemOpen
  \bibfield  {author} {\bibinfo {author} {\bibfnamefont {L.}~\bibnamefont
  {Baudis}}, \bibinfo {author} {\bibfnamefont {G.}~\bibnamefont {Kessler}},
  \bibinfo {author} {\bibfnamefont {P.}~\bibnamefont {Klos}}, \bibinfo {author}
  {\bibfnamefont {R.~F.}\ \bibnamefont {Lang}}, \bibinfo {author}
  {\bibfnamefont {J.}~\bibnamefont {Men\'endez}}, \bibinfo {author}
  {\bibfnamefont {S.}~\bibnamefont {Reichard}}, \ and\ \bibinfo {author}
  {\bibfnamefont {A.}~\bibnamefont {Schwenk}},\ }\href {\doibase
  10.1103/PhysRevD.88.115014} {\bibfield  {journal} {\bibinfo  {journal} {Phys.
  Rev. D}\ }\textbf {\bibinfo {volume} {88}},\ \bibinfo {pages} {115014}
  (\bibinfo {year} {2013})}\BibitemShut {NoStop}%
\bibitem [{\citenamefont {Gazda}\ \emph {et~al.}(2017)\citenamefont {Gazda},
  \citenamefont {Catena},\ and\ \citenamefont
  {Forss\'en}}]{PhysRevD.95.103011}%
  \BibitemOpen
  \bibfield  {author} {\bibinfo {author} {\bibfnamefont {D.}~\bibnamefont
  {Gazda}}, \bibinfo {author} {\bibfnamefont {R.}~\bibnamefont {Catena}}, \
  and\ \bibinfo {author} {\bibfnamefont {C.}~\bibnamefont {Forss\'en}},\ }\href
  {\doibase 10.1103/PhysRevD.95.103011} {\bibfield  {journal} {\bibinfo
  {journal} {Phys. Rev. D}\ }\textbf {\bibinfo {volume} {95}},\ \bibinfo
  {pages} {103011} (\bibinfo {year} {2017})}\BibitemShut {NoStop}%
\bibitem [{\citenamefont {Hoferichter}\ \emph {et~al.}(2019)\citenamefont
  {Hoferichter}, \citenamefont {Klos}, \citenamefont {Men\'endez},\ and\
  \citenamefont {Schwenk}}]{PhysRevD.99.055031}%
  \BibitemOpen
  \bibfield  {author} {\bibinfo {author} {\bibfnamefont {M.}~\bibnamefont
  {Hoferichter}}, \bibinfo {author} {\bibfnamefont {P.}~\bibnamefont {Klos}},
  \bibinfo {author} {\bibfnamefont {J.}~\bibnamefont {Men\'endez}}, \ and\
  \bibinfo {author} {\bibfnamefont {A.}~\bibnamefont {Schwenk}},\ }\href
  {\doibase 10.1103/PhysRevD.99.055031} {\bibfield  {journal} {\bibinfo
  {journal} {Phys. Rev. D}\ }\textbf {\bibinfo {volume} {99}},\ \bibinfo
  {pages} {055031} (\bibinfo {year} {2019})}\BibitemShut {NoStop}%
\bibitem [{\citenamefont {Lloyd}(1996)}]{lloddse}%
  \BibitemOpen
  \bibfield  {author} {\bibinfo {author} {\bibfnamefont {S.}~\bibnamefont
  {Lloyd}},\ }\href@noop {} {\bibfield  {journal} {\bibinfo  {journal}
  {Science}\ }\textbf {\bibinfo {volume} {273}},\ \bibinfo {pages} {1073}
  (\bibinfo {year} {1996})}\BibitemShut {NoStop}%
\bibitem [{\citenamefont {Clemente}\ \emph {et~al.}(2020)\citenamefont
  {Clemente}, \citenamefont {Cardinali}, \citenamefont {Bonati}, \citenamefont
  {Calore}, \citenamefont {Cosmai}, \citenamefont {D'Elia}, \citenamefont
  {Gabbana}, \citenamefont {Rossini}, \citenamefont {Schifano}, \citenamefont
  {Tripiccione},\ and\ \citenamefont {Vadacchino}}]{PhysRevD.101.074510}%
  \BibitemOpen
  \bibfield  {author} {\bibinfo {author} {\bibfnamefont {G.}~\bibnamefont
  {Clemente}}, \bibinfo {author} {\bibfnamefont {M.}~\bibnamefont {Cardinali}},
  \bibinfo {author} {\bibfnamefont {C.}~\bibnamefont {Bonati}}, \bibinfo
  {author} {\bibfnamefont {E.}~\bibnamefont {Calore}}, \bibinfo {author}
  {\bibfnamefont {L.}~\bibnamefont {Cosmai}}, \bibinfo {author} {\bibfnamefont
  {M.}~\bibnamefont {D'Elia}}, \bibinfo {author} {\bibfnamefont
  {A.}~\bibnamefont {Gabbana}}, \bibinfo {author} {\bibfnamefont
  {D.}~\bibnamefont {Rossini}}, \bibinfo {author} {\bibfnamefont {F.~S.}\
  \bibnamefont {Schifano}}, \bibinfo {author} {\bibfnamefont {R.}~\bibnamefont
  {Tripiccione}}, \ and\ \bibinfo {author} {\bibfnamefont {D.}~\bibnamefont
  {Vadacchino}} (\bibinfo {collaboration} {QuBiPF Collaboration}),\ }\href
  {\doibase 10.1103/PhysRevD.101.074510} {\bibfield  {journal} {\bibinfo
  {journal} {Phys. Rev. D}\ }\textbf {\bibinfo {volume} {101}},\ \bibinfo
  {pages} {074510} (\bibinfo {year} {2020})}\BibitemShut {NoStop}%
\bibitem [{\citenamefont {Loh}\ \emph {et~al.}(1990)\citenamefont {Loh},
  \citenamefont {Gubernatis}, \citenamefont {Scalettar}, \citenamefont {White},
  \citenamefont {Scalapino},\ and\ \citenamefont {Sugar}}]{Loh:1990zz}%
  \BibitemOpen
  \bibfield  {author} {\bibinfo {author} {\bibfnamefont {E.~Y.}\ \bibnamefont
  {Loh}, \bibfnamefont {Jr.}}, \bibinfo {author} {\bibfnamefont {J.~E.}\
  \bibnamefont {Gubernatis}}, \bibinfo {author} {\bibfnamefont {R.~T.}\
  \bibnamefont {Scalettar}}, \bibinfo {author} {\bibfnamefont {S.~R.}\
  \bibnamefont {White}}, \bibinfo {author} {\bibfnamefont {D.~J.}\ \bibnamefont
  {Scalapino}}, \ and\ \bibinfo {author} {\bibfnamefont {R.~L.}\ \bibnamefont
  {Sugar}},\ }\href@noop {} {\bibfield  {journal} {\bibinfo  {journal}
  {Physical Review B}\ }\textbf {\bibinfo {volume} {41}},\ \bibinfo {pages}
  {9301} (\bibinfo {year} {1990})}\BibitemShut {NoStop}%
\bibitem [{\citenamefont {Krieger}(1977)}]{krieger1977comparison}%
  \BibitemOpen
  \bibfield  {author} {\bibinfo {author} {\bibfnamefont {S.}~\bibnamefont
  {Krieger}},\ }\href@noop {} {\bibfield  {journal} {\bibinfo  {journal}
  {Nuclear Physics A}\ }\textbf {\bibinfo {volume} {276}},\ \bibinfo {pages}
  {12} (\bibinfo {year} {1977})}\BibitemShut {NoStop}%
\bibitem [{\citenamefont {Hoodbhoy}\ and\ \citenamefont
  {Negele}(1978)}]{PhysRevC.18.2380}%
  \BibitemOpen
  \bibfield  {author} {\bibinfo {author} {\bibfnamefont {P.}~\bibnamefont
  {Hoodbhoy}}\ and\ \bibinfo {author} {\bibfnamefont {J.~W.}\ \bibnamefont
  {Negele}},\ }\href {\doibase 10.1103/PhysRevC.18.2380} {\bibfield  {journal}
  {\bibinfo  {journal} {Phys. Rev. C}\ }\textbf {\bibinfo {volume} {18}},\
  \bibinfo {pages} {2380} (\bibinfo {year} {1978})}\BibitemShut {NoStop}%
\bibitem [{\citenamefont {Wahlen-Strothman}\ \emph {et~al.}(2017)\citenamefont
  {Wahlen-Strothman}, \citenamefont {Henderson}, \citenamefont {Hermes},
  \citenamefont {Degroote}, \citenamefont {Qiu}, \citenamefont {Zhao},
  \citenamefont {Dukelsky},\ and\ \citenamefont
  {Scuseria}}]{wahlen2017merging}%
  \BibitemOpen
  \bibfield  {author} {\bibinfo {author} {\bibfnamefont {J.~M.}\ \bibnamefont
  {Wahlen-Strothman}}, \bibinfo {author} {\bibfnamefont {T.~M.}\ \bibnamefont
  {Henderson}}, \bibinfo {author} {\bibfnamefont {M.~R.}\ \bibnamefont
  {Hermes}}, \bibinfo {author} {\bibfnamefont {M.}~\bibnamefont {Degroote}},
  \bibinfo {author} {\bibfnamefont {Y.}~\bibnamefont {Qiu}}, \bibinfo {author}
  {\bibfnamefont {J.}~\bibnamefont {Zhao}}, \bibinfo {author} {\bibfnamefont
  {J.}~\bibnamefont {Dukelsky}}, \ and\ \bibinfo {author} {\bibfnamefont
  {G.~E.}\ \bibnamefont {Scuseria}},\ }\href@noop {} {\bibfield  {journal}
  {\bibinfo  {journal} {The Journal of Chemical Physics}\ }\textbf {\bibinfo
  {volume} {146}},\ \bibinfo {pages} {054110} (\bibinfo {year}
  {2017})}\BibitemShut {NoStop}%
\bibitem [{\citenamefont {Stoica}\ \emph {et~al.}(2001)\citenamefont {Stoica},
  \citenamefont {Mihut},\ and\ \citenamefont {Suhonen}}]{PhysRevC.64.017303}%
  \BibitemOpen
  \bibfield  {author} {\bibinfo {author} {\bibfnamefont {S.}~\bibnamefont
  {Stoica}}, \bibinfo {author} {\bibfnamefont {I.}~\bibnamefont {Mihut}}, \
  and\ \bibinfo {author} {\bibfnamefont {J.}~\bibnamefont {Suhonen}},\ }\href
  {\doibase 10.1103/PhysRevC.64.017303} {\bibfield  {journal} {\bibinfo
  {journal} {Phys. Rev. C}\ }\textbf {\bibinfo {volume} {64}},\ \bibinfo
  {pages} {017303} (\bibinfo {year} {2001})}\BibitemShut {NoStop}%
\bibitem [{\citenamefont {Severyukhin}\ \emph {et~al.}(2006)\citenamefont
  {Severyukhin}, \citenamefont {Bender},\ and\ \citenamefont
  {Heenen}}]{PhysRevC.74.024311}%
  \BibitemOpen
  \bibfield  {author} {\bibinfo {author} {\bibfnamefont {A.~P.}\ \bibnamefont
  {Severyukhin}}, \bibinfo {author} {\bibfnamefont {M.}~\bibnamefont {Bender}},
  \ and\ \bibinfo {author} {\bibfnamefont {P.-H.}\ \bibnamefont {Heenen}},\
  }\href {\doibase 10.1103/PhysRevC.74.024311} {\bibfield  {journal} {\bibinfo
  {journal} {Phys. Rev. C}\ }\textbf {\bibinfo {volume} {74}},\ \bibinfo
  {pages} {024311} (\bibinfo {year} {2006})}\BibitemShut {NoStop}%
\bibitem [{\citenamefont {Lacroix}(2009)}]{PhysRevC.79.014301}%
  \BibitemOpen
  \bibfield  {author} {\bibinfo {author} {\bibfnamefont {D.}~\bibnamefont
  {Lacroix}},\ }\href {\doibase 10.1103/PhysRevC.79.014301} {\bibfield
  {journal} {\bibinfo  {journal} {Phys. Rev. C}\ }\textbf {\bibinfo {volume}
  {79}},\ \bibinfo {pages} {014301} (\bibinfo {year} {2009})}\BibitemShut
  {NoStop}%
\bibitem [{\citenamefont {Bertolli}\ and\ \citenamefont
  {Papenbrock}(2008)}]{PhysRevC.78.064310}%
  \BibitemOpen
  \bibfield  {author} {\bibinfo {author} {\bibfnamefont {M.~G.}\ \bibnamefont
  {Bertolli}}\ and\ \bibinfo {author} {\bibfnamefont {T.}~\bibnamefont
  {Papenbrock}},\ }\href {\doibase 10.1103/PhysRevC.78.064310} {\bibfield
  {journal} {\bibinfo  {journal} {Phys. Rev. C}\ }\textbf {\bibinfo {volume}
  {78}},\ \bibinfo {pages} {064310} (\bibinfo {year} {2008})}\BibitemShut
  {NoStop}%
\bibitem [{\citenamefont {Plesch}\ and\ \citenamefont
  {Brukner}(2011)}]{quantumstateprep}%
  \BibitemOpen
  \bibfield  {author} {\bibinfo {author} {\bibfnamefont {M.}~\bibnamefont
  {Plesch}}\ and\ \bibinfo {author} {\bibfnamefont {C.}~\bibnamefont
  {Brukner}},\ }\href {\doibase 10.1103/PhysRevA.83.032302} {\bibfield
  {journal} {\bibinfo  {journal} {Phys. Rev. A}\ }\textbf {\bibinfo {volume}
  {83}},\ \bibinfo {pages} {032302} (\bibinfo {year} {2011})}\BibitemShut
  {NoStop}%
\bibitem [{\citenamefont {Ortiz}\ \emph {et~al.}(2005)\citenamefont {Ortiz},
  \citenamefont {Somma}, \citenamefont {Dukelsky},\ and\ \citenamefont
  {Rombouts}}]{ORTIZ2005421}%
  \BibitemOpen
  \bibfield  {author} {\bibinfo {author} {\bibfnamefont {G.}~\bibnamefont
  {Ortiz}}, \bibinfo {author} {\bibfnamefont {R.}~\bibnamefont {Somma}},
  \bibinfo {author} {\bibfnamefont {J.}~\bibnamefont {Dukelsky}}, \ and\
  \bibinfo {author} {\bibfnamefont {S.}~\bibnamefont {Rombouts}},\ }\href
  {\doibase https://doi.org/10.1016/j.nuclphysb.2004.11.008} {\bibfield
  {journal} {\bibinfo  {journal} {Nuclear Physics B}\ }\textbf {\bibinfo
  {volume} {707}},\ \bibinfo {pages} {421 } (\bibinfo {year}
  {2005})}\BibitemShut {NoStop}%
\bibitem [{\citenamefont {Lerma~H.}\ and\ \citenamefont
  {Dukelsky}(2013)}]{LermaH.:2013cla}%
  \BibitemOpen
  \bibfield  {author} {\bibinfo {author} {\bibfnamefont {S.}~\bibnamefont
  {Lerma~H.}}\ and\ \bibinfo {author} {\bibfnamefont {J.}~\bibnamefont
  {Dukelsky}},\ }\href {\doibase 10.1016/j.nuclphysb.2013.01.019} {\bibfield
  {journal} {\bibinfo  {journal} {Nucl. Phys.}\ }\textbf {\bibinfo {volume}
  {B870}},\ \bibinfo {pages} {421} (\bibinfo {year} {2013})}\BibitemShut
  {NoStop}%
\bibitem [{\citenamefont {Somma}(2005)}]{somma2005quantum}%
  \BibitemOpen
  \bibfield  {author} {\bibinfo {author} {\bibfnamefont {R.~D.}\ \bibnamefont
  {Somma}},\ }\emph {\bibinfo {title} {Quantum computation, complexity, and
  many-body physics}},\ \href@noop {} {Ph.D. thesis} (\bibinfo {year} {2005}),\
  \bibinfo {note} {{Ph.D.} Thesis}\BibitemShut {NoStop}%
\bibitem [{\citenamefont {Robbins}\ and\ \citenamefont
  {Love}()}]{robbins2021benchmarking}%
  \BibitemOpen
  \bibfield  {author} {\bibinfo {author} {\bibfnamefont {K.}~\bibnamefont
  {Robbins}}\ and\ \bibinfo {author} {\bibfnamefont {P.~J.}\ \bibnamefont
  {Love}},\ }\href@noop {} {\ }\Eprint {http://arxiv.org/abs/2105.06761}
  {arXiv:2105.06761} \BibitemShut {NoStop}%
\bibitem [{\citenamefont {Rubio}\ \emph {et~al.}(2019)\citenamefont {Rubio},
  \citenamefont {Bevenius}, \citenamefont {Costa~Hamido}, \citenamefont
  {Carballo}, \citenamefont {Rodríguez~Davila}, \citenamefont {Cruz-Benito},\
  and\ \citenamefont {Hu}}]{qiskitAbbrev}%
  \BibitemOpen
  \bibfield  {author} {\bibinfo {author} {\bibfnamefont {J.}~\bibnamefont
  {Rubio}}, \bibinfo {author} {\bibfnamefont {D.}~\bibnamefont {Bevenius}},
  \bibinfo {author} {\bibfnamefont {O.}~\bibnamefont {Costa~Hamido}}, \bibinfo
  {author} {\bibfnamefont {J.}~\bibnamefont {Carballo}}, \bibinfo {author}
  {\bibfnamefont {A.}~\bibnamefont {Rodríguez~Davila}}, \bibinfo {author}
  {\bibfnamefont {J.}~\bibnamefont {Cruz-Benito}}, \ and\ \bibinfo {author}
  {\bibfnamefont {W.}~\bibnamefont {Hu}},\ }\href
  {\url{https://github.com/qiskit-community/qiskit-js#readme}} {\enquote
  {\bibinfo {title} {{Qiskit (Quantum Information Science Kit) for
  JavaScript}},}\ } (\bibinfo {year} {2019})\BibitemShut {NoStop}%
\bibitem [{\citenamefont {{Qiskit Development
  Team}}(2020{\natexlab{a}})}]{ibmGlossary}%
  \BibitemOpen
  \bibfield  {author} {\bibinfo {author} {\bibnamefont {{Qiskit Development
  Team}}},\ }\href@noop {} {\enquote {\bibinfo {title} {{IBM Quantum
  Experience} glossary},}\ }\bibinfo {howpublished}
  {\url{https://quantum-computing.ibm.com/docs/glossary/}} (\bibinfo {year}
  {2020}{\natexlab{a}}),\ \bibinfo {note} {accessed: 2020-10-08}\BibitemShut
  {NoStop}%
\bibitem [{\citenamefont {{IBM Quantum}}(2020)}]{OpenQASM}%
  \BibitemOpen
  \bibfield  {author} {\bibinfo {author} {\bibnamefont {{IBM Quantum}}},\
  }\href@noop {} {\enquote {\bibinfo {title} {{IBM Quantum Experience Open QASM
  Simulator}},}\ }\bibinfo {howpublished}
  {\url{https://quantum-computing.ibm.com/docs/manage/backends/simulators/}}
  (\bibinfo {year} {2020}),\ \bibinfo {note} {accessed: 2020-10-08}\BibitemShut
  {NoStop}%
\bibitem [{\citenamefont {{Qiskit Development
  Team}}(2020{\natexlab{b}})}]{GeneralNoise}%
  \BibitemOpen
  \bibfield  {author} {\bibinfo {author} {\bibnamefont {{Qiskit Development
  Team}}},\ }\href@noop {} {\enquote {\bibinfo {title} {Qiskit: Device backend
  noise model simulations},}\ }\bibinfo {howpublished}
  {\url{https://qiskit.org/documentation/tutorials/simulators/2_device_noise_simulation.html}}
  (\bibinfo {year} {2020}{\natexlab{b}}),\ \bibinfo {note} {accessed:
  2020-10-09}\BibitemShut {NoStop}%
\bibitem [{\citenamefont {{Qiskit Development
  Team}}(2020{\natexlab{c}})}]{ReadOutNoise}%
  \BibitemOpen
  \bibfield  {author} {\bibinfo {author} {\bibnamefont {{Qiskit Development
  Team}}},\ }\href@noop {} {\enquote {\bibinfo {title} {Qiskit:
  Readouterror},}\ }\bibinfo {howpublished}
  {\url{https://qiskit.org/documentation/stubs/qiskit.providers.aer.noise.ReadoutError.html}}
  (\bibinfo {year} {2020}{\natexlab{c}}),\ \bibinfo {note} {accessed:
  2020-10-09}\BibitemShut {NoStop}%
\bibitem [{\citenamefont {{Qiskit Development
  Team}}(2020{\natexlab{d}})}]{BackendProperties}%
  \BibitemOpen
  \bibfield  {author} {\bibinfo {author} {\bibnamefont {{Qiskit Development
  Team}}},\ }\href@noop {} {\enquote {\bibinfo {title} {Qiskit:
  Ibmqbackend.properties},}\ }\bibinfo {howpublished}
  {\url{https://qiskit.org/documentation/stubs/qiskit.providers.ibmq.IBMQBackend.properties.html}}
  (\bibinfo {year} {2020}{\natexlab{d}}),\ \bibinfo {note} {accessed:
  2020-10-09}\BibitemShut {NoStop}%
\bibitem [{\citenamefont {{Qiskit Development
  Team}}(2020{\natexlab{e}})}]{U1gate}%
  \BibitemOpen
  \bibfield  {author} {\bibinfo {author} {\bibnamefont {{Qiskit Development
  Team}}},\ }\href@noop {} {\enquote {\bibinfo {title} {Qiskit: U1gate},}\
  }\bibinfo {howpublished}
  {\url{https://qiskit.org/documentation/stubs/qiskit.circuit.library.U1Gate.html}}
  (\bibinfo {year} {2020}{\natexlab{e}}),\ \bibinfo {note} {accessed:
  2020-10-27}\BibitemShut {NoStop}%
\bibitem [{\citenamefont {{Qiskit Development
  Team}}(2020{\natexlab{f}})}]{U3gate}%
  \BibitemOpen
  \bibfield  {author} {\bibinfo {author} {\bibnamefont {{Qiskit Development
  Team}}},\ }\href@noop {} {\enquote {\bibinfo {title} {Qiskit: U3gate},}\
  }\bibinfo {howpublished}
  {\url{https://qiskit.org/documentation/stubs/qiskit.circuit.library.U3Gate.html}}
  (\bibinfo {year} {2020}{\natexlab{f}}),\ \bibinfo {note} {accessed:
  2020-10-27}\BibitemShut {NoStop}%
\bibitem [{\citenamefont {Nation}(2020)}]{QiskitBenchmarking}%
  \BibitemOpen
  \bibfield  {author} {\bibinfo {author} {\bibfnamefont {P.}~\bibnamefont
  {Nation}},\ }\href@noop {} {\enquote {\bibinfo {title} {Randomized
  benchmarking},}\ }\bibinfo {howpublished}
  {\url{https://github.com/Qiskit/qiskit-tutorials/blob/master/tutorials/noise/4_randomized_benchmarking.ipynb}}
  (\bibinfo {year} {2020})\BibitemShut {NoStop}%
\bibitem [{\citenamefont {{Qiskit Development
  Team}}(2020{\natexlab{g}})}]{NoiseModelApprox}%
  \BibitemOpen
  \bibfield  {author} {\bibinfo {author} {\bibnamefont {{Qiskit Development
  Team}}},\ }\href@noop {} {\enquote {\bibinfo {title} {Qiskit: Noise
  models},}\ }\bibinfo {howpublished}
  {\url{https://qiskit.org/documentation/apidoc/aer_noise.html}} (\bibinfo
  {year} {2020}{\natexlab{g}}),\ \bibinfo {note} {accessed:
  2020-10-09}\BibitemShut {NoStop}%
\bibitem [{\citenamefont {{Qiskit Development
  Team}}(2020{\natexlab{h}})}]{QiskitErrorMitigate}%
  \BibitemOpen
  \bibfield  {author} {\bibinfo {author} {\bibnamefont {{Qiskit Development
  Team}}},\ }\href@noop {} {\enquote {\bibinfo {title} {Qiskit: Measurement
  error mitigation},}\ }\bibinfo {howpublished}
  {\url{https://qiskit.org/textbook/ch-quantum-hardware/measurement-error-mitigation.html}}
  (\bibinfo {year} {2020}{\natexlab{h}}),\ \bibinfo {note} {accessed:
  2020-10-09}\BibitemShut {NoStop}%
\bibitem [{\citenamefont {Satzinger}\ \emph {et~al.}(2021)\citenamefont
  {Satzinger}, \citenamefont {Liu}, \citenamefont {Smith}, \citenamefont
  {Knapp}, \citenamefont {Newman}, \citenamefont {Jones}, \citenamefont {Chen},
  \citenamefont {Quintana}, \citenamefont {Mi}, \citenamefont {Dunsworth},
  \citenamefont {Gidney}, \citenamefont {Aleiner}, \citenamefont {Arute},
  \citenamefont {Arya}, \citenamefont {Atalaya}, \citenamefont {Babbush},
  \citenamefont {Bardin}, \citenamefont {Barends}, \citenamefont {Basso},
  \citenamefont {Bengtsson} \emph {et~al.}}]{satzinger2021realizing}%
  \BibitemOpen
  \bibfield  {author} {\bibinfo {author} {\bibfnamefont {K.~J.}\ \bibnamefont
  {Satzinger}}, \bibinfo {author} {\bibfnamefont {Y.}~\bibnamefont {Liu}},
  \bibinfo {author} {\bibfnamefont {A.}~\bibnamefont {Smith}}, \bibinfo
  {author} {\bibfnamefont {C.}~\bibnamefont {Knapp}}, \bibinfo {author}
  {\bibfnamefont {M.}~\bibnamefont {Newman}}, \bibinfo {author} {\bibfnamefont
  {C.}~\bibnamefont {Jones}}, \bibinfo {author} {\bibfnamefont
  {Z.}~\bibnamefont {Chen}}, \bibinfo {author} {\bibfnamefont {C.}~\bibnamefont
  {Quintana}}, \bibinfo {author} {\bibfnamefont {X.}~\bibnamefont {Mi}},
  \bibinfo {author} {\bibfnamefont {A.}~\bibnamefont {Dunsworth}}, \bibinfo
  {author} {\bibfnamefont {C.}~\bibnamefont {Gidney}}, \bibinfo {author}
  {\bibfnamefont {I.}~\bibnamefont {Aleiner}}, \bibinfo {author} {\bibfnamefont
  {F.}~\bibnamefont {Arute}}, \bibinfo {author} {\bibfnamefont
  {K.}~\bibnamefont {Arya}}, \bibinfo {author} {\bibfnamefont {J.}~\bibnamefont
  {Atalaya}}, \bibinfo {author} {\bibfnamefont {R.}~\bibnamefont {Babbush}},
  \bibinfo {author} {\bibfnamefont {J.~C.}\ \bibnamefont {Bardin}}, \bibinfo
  {author} {\bibfnamefont {R.}~\bibnamefont {Barends}}, \bibinfo {author}
  {\bibfnamefont {J.}~\bibnamefont {Basso}}, \bibinfo {author} {\bibfnamefont
  {A.}~\bibnamefont {Bengtsson}},  \emph {et~al.},\ }\href@noop {} {\enquote
  {\bibinfo {title} {Realizing topologically ordered states on a quantum
  processor},}\ } (\bibinfo {year} {2021}),\ \Eprint
  {http://arxiv.org/abs/2104.01180} {arXiv:2104.01180 [quant-ph]} \BibitemShut
  {NoStop}%
\end{thebibliography}%
\bibliographystyle{apsrev4-1}

\end{document}